\documentclass[twocolumn]{aastex63}
\usepackage{tikz}
\usepackage{pgfplots}
\pgfplotsset{compat=1.14}
\usetikzlibrary{intersections}
\usepackage{booktabs}
\newcommand{\ra}[1]{\renewcommand{\arraystretch}{#1}}

\newcommand{\source}{{1ES 1959+650}}
\newcommand{\astrosat}{{\it AstroSat}}
\newcommand{\swift}{{\it Swift}}
\newcommand{\fermi}{{\it Fermi}-LAT}
\newcommand{\xmm}{{\it XMM-Newton}}
\newcommand{\miro}{Mt. Abu Infrared Observatory, India (MIRO)}

\received{\today}
\submitjournal{ApJ}

\shorttitle{X-ray activities in 1ES 1959+650}
\shortauthors{Chandra S., et al.}

\begin{document}

\title{X-ray Observations of 1ES 1959+650 in its high activity state in 2016-2017 with AstroSat and Swift}

\correspondingauthor{Sunil Chandra}
\email{sunil.chandra355@gmail.com, sunil.chandra@nwu.ac.in}

\author[0000-0002-8776-1835]{Sunil Chandra}
\affiliation{Center for Space Research,
North-West University, Potchefstroom,
2520, South Africa}

\author[0000-0002-8434-5692]{Markus Boettcher}
\affiliation{Center for Space Research,
North-West University, Potchefstroom,
2520, South Africa}

\author[0000-0001-5430-4374]{Pranjupriya Goswami}
\affiliation{Department of Physics, Tezpur University, Assam, India}

\author[0000-0001-6952-3887]{Kulinder Pal Singh}
\affiliation{Indian Institute of Science Education and Research, Mohali, Punjab, India}
\affiliation{Department of Astronomy and Astrophysics, Tata Institute of Fundamental
Research, 1 Homi Bhabha Road, Mumbai 400005, India}

\author[0000-0001-5801-3945]{Michael Zacharias}
\affiliation{Laboratoire Univers et Théories, Observatoire de Paris, Université PSL, CNRS, Université de Paris, 92190 Meudon, France}
\affiliation{Center for Space Research, 
North-West University, Potchefstroom,
2520, South Africa}

\author[0000-0002-7862-1056]{Navpreet Kaur}
\affiliation{KTH Royal Institute of Technology, Sweden}

\author[0000-0002-6351-5808]{Sudip Bhattacharyya}
\affiliation{Department of Astronomy and Astrophysics, Tata Institute of Fundamental
Research, 1 Homi Bhabha Road, Mumbai 400005, India}

\author[0000-0002-7721-3827]{Shashikiran Ganesh}
\affiliation{Physical Research Laboratory Ahmedabad, India}

\author[0000-0001-8823-479X]{Daniela Dorner}
\affiliation{Department of Physics and Astronomy, Julius Maximilian University Würzburg, Am Hubland, 97074 Würzburg, Germany}
\begin{abstract}
We present a comprehensive multi-frequency study of the HBL \source\ using data from various facilities during the period 2016-2017, including X-ray data from \astrosat\ and {\it Swift} during the historically high X-ray flux state of the source observed until February 2021. The unprecedented quality of X-ray data from high cadence monitoring with the \astrosat\ during 2016-2017 enables us to establish a detailed description of X-ray flares in \source. The synchrotron peak shifts significantly between different flux states, in a manner consistent with a geometric (changing Doppler factor) interpretation. A time-dependent leptonic diffusive-shock-acceleration and radiation transfer model is used to reproduce the spectral energy distributions (SEDs) and X-ray light curves, to provide insight into the particle acceleration during the major activity periods observed in 2016 and 2017. 
The extensive data of \swift-XRT from December 2015 to February 2021 (Exp. = 411.3 ks) reveals a positive correlation between flux and peak position.  
\end{abstract}

\keywords{(galaxies), BL Lac Objects --- 
HBLs --- Individual (1ES 1959+650)}

\section{Introduction}\label{sec:intro}

 Blazars are a subclass of active galactic nuclei (AGN) with a jet of relativistic plasma streaming along or very close to the line of sight 
\citep{Blandford+1978}. The observed electromagnetic (EM) emission, which is predominantly non-thermal in nature, is considered to be emanating from the relativistic jet. The observed features of blazars include superluminal motion of radio-jet components, high optical polarisation and strong continuum emission variable at time scales ranging from a few minutes to years, across the entire EM spectrum.

The broadband spectral energy distribution (SED; $\nu$  v/s  $\nu$F$_\nu$ plot) of blazars consists of two distinct broad continuum hump like structures with the first one peaking somewhere in sub-mm to soft X-rays, whereas the second one peaks at MeV to TeV energies  \citep{Urray+1995}. The low-energy component of the SED is mostly due to the synchrotron emission from relativistic electrons/positrons gyrating around the magnetic field in the relativistic jet. This emission component, in some cases, is superimposed by significant thermal contributions from, e.g., the accretion disk, a hot corona accompanying the accretion disk, and/or an obscuring dusty torus. On the other hand, the physical mechanisms behind high energy emission (MeV to TeV) are not well established and two families of models namely, a) leptonic models, and, b) hadronic models, both appear to be viable mechanisms to explain the X-ray through $\gamma$-ray emission. In hadronic models the high energy emission is produced by relativistic protons through proton synchrotron radiation and photo-pion production, followed by pion decay and electromagnetic cascades \citep[e.g.,][]{1992A&A...253L..21M, 1993PhRvD..47.5270N, 1993PhRvD..48.2408M, 2001AIPC..558..700P, 2003APh....18..593M, 2013ApJ...768...54B}. Leptonic models assume the jet protons to be cold enough not to contribute to the radiative output, and high-energy emission is produced by inverse Compton scattering of low energy seed photons by the ultra-relativistic leptons (e$^-$/e$^+$). 
The seed photons may come from the synchrotron radiation field in the emission region, which are  up-scattered by the same leptons that produced the synchrotron radiation (Synchrotron Self-Compton = SSC) \citep{Ghisellini+1989,Bloom1996,1997A&A...320...19M}). Alternatively, if the seed photons originate external to the emission region (e.g., from the accretion-disk, the dusty torus, or the broad-line region) then the process is termed as External Compton (EC). \citep{1992A&A...256L..27D,1998MNRAS.301..451G}). 

 \source\ \citep[$z = 0.048$;][]{1996ApJS..104..251P} is a prominent high-synchrotron-peaked blazar. It was first detected in X-rays during the Slew Survey with the \emph{Einstein} Imaging Proportional Counter (IPC) \citep{1992ApJS...80..257E}, followed by \emph{BeppoSAX} \citep{2002A&A...383..410B}, \emph{RXTE}, \emph{Swift}, \emph{XMM-Newton} \citep{Tagliaferri2003, Massaro2008} in later years. This source is also a prominent TeV $\gamma$-ray emitter, with the first detection at TeV energies, reported by the Utah Seven-Telescope Array collaboration in 1998 \citep{1999ICRC....3..370N}.

The historical observations establish \source\ to be a High-frequency peaked BL Lac object (HBL) in which the synchrotron peak of the broadband SED appears in UV -- X-ray band \citep{Krawczynski2004,Kapanadze2016b, 2010ApJ...716...30A}. This source exhibits strong flux variability across almost the entire EM spectrum. The flux increase of up to 3-4 orders of magnitude in the optical, X-ray, and TeV energy bands during the short/erratic flares have been witnessed for \source\ \citep{Perlman2005,Krawczynski2004,Kapanadze2016a}.
The rapid variability and its frequency dependence provide crucial insight into the physical processes of particle acceleration and radiation mechanisms as well as the geometry and size of the emission region \citep[e.g.,][]{1995MNRAS.273..583D}. 
 
Recent high-sensitivity X-ray observations have found several high flux states and strong X-ray outbursts of this source. \emph{XMM-Newton} and \emph{RXTE}-PCA observations in 2002 -- 2003 revealed strong X-ray flares with flux variations by a factor up to $\sim$ 4.2 \citep{Perlman2005, Krawczynski2004}. Many of such frequently occurring strong X-ray flares were reported by \cite{Kapanadze2016a} during 2006 -- 2014 using \emph{Swift}-XRT observations. The source underwent a number of active states and an unprecedented X-ray flaring activity during August 2015 -- January 2016 that was observed by \emph{Swift}-XRT. The observed count rate was reported to vary by a factor of $\sim$5.7 with maximum value above 20 cts/s, with simultaneous high flux activity in TeV energy band \citep{Kapanadze2016b,Kaur2017,2018A&A...611A..44P,MAGIC2020}. During this large flare, the synchrotron peak position of the SED showed a tendency to shift towards the higher X-ray energies accompanied by a hard X-ray spectral index. The detailed X-ray spectral studies further confirmed the harder-when-brighter trend \citep{Tagliaferri2003,MAGIC2020}. However, during most of these epochs with an X-ray flare, the TeV counterpart was found to be in low flux states. On the other hand, in several multi-wavelength campaigns,
``orphan" flares in VHE (Very High Energy, used for TeV) $\gamma$-rays (not accompanied by a simultaneous X-ray flare) have been reported in June 2002 \citep{2004ApJ...601..151K} and April -- June 2012 \citep{2014ApJ...797...89A}. 
 
The  uncorrelated variability is inconsistent with the simplest one-zone SSC models, which are often successful in reproducing the broadband emission of HBLs, but have proven to be inadequate to explain several aspects of emission in many studies \citep{Krawczynski2004,2018A&A...611A..44P,MAGIC2020}. ``Orphan'' flares hint at a more complex geometry and/or underlying particle distribution, such as those invoked in multiple-component SSC models and/or external Compton models (e.g., the synchrotron mirror model, \cite{1999ApL&C..39..129B}) within the leptonic schemes. Recently, \citet{shah2021astrosat}, have reported an anti-correlation between the photon index and X-ray flux using a broken power-law
for analysing only a segment of  X-ray data presented here. 

In this work, we present a detailed investigation of the X-ray spectral and light curve features of \source\ observed by \emph{AstroSat} in 2016 -- 2017. Our main focus here is to understand the distinct, irregular X-ray outbursts observed during this period by both the Soft X-Ray Telescope (SXT) \& Large Area Proportional Counter (LAXPC) instruments aboard \emph{AstroSat}.

These data are supplemented with simultaneous/quasi-simultaneous XRT data extending before and after 2016--2017, spanning over 6 years from January 30, 2015 -- February 09, 2021, and also other multi-wavelength data to probe the evolution of the underlying non-thermal particle distribution. In order to consistently fit the SEDs and the light curves obtained during these erratic flares observed with \emph{AstroSat}, we adopt the time-dependent multi-zone shock acceleration and radiation transfer model, as described by \cite{BB19}, to investigate the nature of shocks responsible for the observed spectral variability. We further provide a detailed analysis of the time resolved spectra and light curves and their correlation over the span of $\sim$ 6 years.

The paper is structured as follows. Section \ref{sec:dataana} describes the multi-wavelength observation details and the data analysis procedures. In section \ref{sec:Results}, we provide the results of the timing and spectral analysis and the detailed correlation study. Section \ref{sec:Results} contains the results and the interpretation through the modeling of snap-shot SEDs and light curves. In section \ref{discussSum} we summarize our work followed by a comprehensive discussion. 

 \label{subsec:intro}
\section{Observations and Data Analysis}\label{sec:dataana}
\source\ was observed in campaign mode during 2016 and 2017 at various epochs, representing different flux states, using a number of observing facilities including \astrosat, \swift, and \miro. 
The details of the observing epochs, and the respective total exposure times are mentioned in the Table~\ref{tab:timesegref}.   
PASS8 photon data from \fermi\ are also analysed to study the high energy (GeV) emission. 
The following sub-sections provide the details of the observations and analysis procedures.  

\begin{table}[t]
    \centering
     \caption{{ \footnotesize \label{tab:obslog} Details of the data used for the present study.}}
    {\footnotesize \begin{tabular}{clll}
    \hline
         S.N. & Instrument & Total Exposure & Epoch of Observations \\
         \hline
         1 & {\it Fermi}-LAT & -  &  57037.0 to 58072.0  \\
         2 & {\it AstroSat} & 143.9 ks &  57666.2 - 57666.7 \\
           &                &          &  57708.4 - 57709.2 \\
           &                &          &  57695.9 - 57699.4 \\
           &                &          &  58051.1 - 58052.6 \\
4 & {\it Swift} & 411.2 ks &  57052.1 - 59254.4 \\
         5 & MIRO & ----  & 57690.81-57696.85 \\ 
&      & R band: 8.4 hrs &       " \\
           &      & B band: 0.65 hrs & " \\
           &      & V band: 0.65 hrs & " \\
        6  & FACT & 12.4 days    & 57632.88-57719.83  \\
           &      &              & 57997.88-58098.83 \\
         \hline  
    \end{tabular}}
   {\footnotesize {\bf Note:} MIRO: Mt. Abu Infrared Observatory, Rajasthan, India\\ 
   FACT: First G-APD Cherenkov Telescope, La Palma, Spain \footnote{https://fact-project.org/}}
\end{table}

\begin{table*}
\caption{\label{tab:timesegref}Details of \astrosat\ observations.}
{\footnotesize \begin{tabular}{cccccl}
\hline
{\bf S.N.} & {\bf Date of Observation} & {\bf Time Start/Range} &  {\bf SXT Exp.}    & {\bf LAXPC Exp.}  &  {\bf Time-Seg.}\\
           &    (UTC)                  &  (MJD)               &      (ks) & (ks) \\
\hline
1          & 2015-11-20T05:28:57 & 57346.27 &  2.9 & -  &  PV \\
2          & 2016-10-05T03:49:57 & 57666.2 - 57666.6 &  13.4 & 25.4 & T0 \\

3          & 2016-11-03T20:47:00 & 57695.9 &  82.6 & 147.8 & {\bf F1:} T1, T2, T3\\
           &                &                  &    &            &   T4, T5, T6 \\
           &  &.............& .................&................&................\\
           &                                 & 57695.9 - 57696.4 & 8.6 & 30.0 & T1 \\
           &                                & 57696.5 - 57697.3 & 13.5 & 27.4 & T2 \\
           &                                 & 57697.3 - 57698.3 & 21.6 & 18.7 & T3 \\
           &                                 & 57698.3 - 57699.4 & 24.3 & 17.5 & T4 \\
           &                                 & 57696.5 - 57697.8 & 26.6 & 46.3.3 & T5 \\
           &                                 & 57697.8 - 57699.0 & 25.6 & 17.4 & T6 \\
           &  &.............& .................&................&................\\

4          & 2016-11-16T10:30:11 & 57708.4 - 57709.1  &  16.0 & 33.3 & T7 \\
5          & 2017-10-25T02:28:50 & 58051.1  &  35.1 & 85.7 & {\bf F2:} T8, T9, T10 \\
           &  &.............& .................&................&................\\
           &                       & 58051.1 - 58051.7 & 14.15 & 38.7 & T8 \\
           &                       & 58051.7 - 58052.3 & 11.0 & 25.7 & T9  \\
           &                       & 58052.3 - 58052.6 & 9.9 & 16.7  & T10  \\
           &  &.............& .................&................&................ \\
\hline \\
\end{tabular}} \\
Note: The small time segments T1-T6 and T8-T10 are used to generate time-resolved spectra to understand the various phases of flaring activities in 2016 and 2017. The time segment PV represents the first target of opportunity (ToO) observations made in 2015.
\end{table*}
 
\subsection{ {\it Fermi}-LAT }
The PASS8 (P8R3) \fermi\ photon data and corresponding spacecraft data from the beginning of November 2015 to the end of December 2017 are downloaded from the LAT data center\footnote{https://fermi.gsfc.nasa.gov/cgi-bin/ssc/LAT/LATDataQuery.cgi} with a search radius of 30 degree and in an energy range of 30 MeV to 500 GeV. The Fermitools package (version 1.2.1 conda-release) distributed by the Fermi Science Support Center, installed with the most recent release of point source (4FGL) and extended source catalogs, are used to analyse the data. The python package, {\tt fermipy}\footnote{http://fermipy-readthedocs.io/en/latest/}\citep{2017ICRC...35..824W} is used, which facilitates handy wrappers for various procedures of LAT data analysis, as described by the instrument teams, including model optimization, the localization, sanity checks and product extractions etc. The initial selection of parameters includes a bin size of 0.1 pixels for map creation, a zenith angle of accepted events of 90$^\circ$ to exclude or eliminate most of the contamination from secondary $\gamma$-rays contributed by Earth's limb, an energy range 100 MeV - 500 GeV, event type 3 and event class 128. The P8R3\_SOURCE\_V2 instrument response functions (IRFs) are used. The initial source model (XML) is created by including all the point-like and extended sources located within 25$^\circ$ radius of the location of \source\ as listed in the Fourth Fermi-LAT Source catalog \citep[4FGL][]{2020ApJS..247...33A}, as well as the Galactic diffuse (gll\_iem\_v07.fits) and isotropic background emission (iso\_P8R3\_SOURCE\_V2\_v1.txt). The source model for \source\ imported from the 4FGL catalog is ``LogParabola"; however, due to poor photon statistics for the duration of interest to this work, a re-optimization of the source model is performed after forcing the spectral shape of \source\ to ``PowerLaw". The spectral parameters of sources within 5$^\circ$ of \source\ are kept variable while others are kept frozen to their best fit values from the catalog. TS (Test Statistics) maps and diffuse maps are generated to look for any possible GeV source (point and/or diffuse) not included in our model, but none are found. 
Once the model is optimized, the best fit spectral parameters for the GeV part of the SED are estimated using the {\tt sed} procedure of {\tt fermipy} with 2 spectral points per decade in energies. The {\tt lightcurve} procedure of the {\tt fermipy} package is used for generating light curves with 1 day, 2 day and 3 day binning.

The SED and light curve data points with TS $\geq$ 9 (equivalent to $\geq$~3~$\sigma$ significance) and TS $\geq$ 25 ($\geq$~5~$\sigma$) are used for spectral and temporal studies. For lower-significance points, 95~\% upper limits are shown.

\subsection{X-ray Data Analysis}
The data from extensive monitoring over the course of two years using \astrosat\ is used for this study. Complementary data from the Neil Gehrels Swift Observatory is also used for various epochs. Table \ref{tab:swtspec_par} lists the details of the data used for this work. 
The general FTools and several mission specifics tools distributed as part of the heasoft package (version 6.25) and the most recent calibration database\footnote{https://heasarc.gsfc.nasa.gov/docs/heasarc/caldb/} are utilized as appropriately to analyse data from various facilities.      

\subsubsection{{\it AstroSat}-Soft X-ray Telescope (SXT)} 
The SXT aboard {\it AstroSat} is a 2-m approximate Wolter-I type focusing instrument sensitive mainly in 0.3-7.1 keV energy band \citep[][]{2014SPIE.9144E..1SS, 2016SPIE.9905E..1ES, 2017JApA...38...29S}. Its camera assembly uses an e2v CCD, identical to the one flown with {\xmm}-MOS and {\swift}-XRT,  at its focal plane as the main detector system. The observations were carried out in photon counting mode. The source was observed throughout all the satellite orbits when the SXT was pointed at it, taking care that the Sun avoidance angle is$\geq$ 45 degrees and the RAM angle (the angle between the payload axis to the velocity vector direction of the spacecraft) $>$ 12 degrees to ensure the safety of the mirrors and the detector.

Level-2 data provided by the SXT payload operation center (POC) in Mumbai, India, are reduced using the most recent pipeline and calibration database (version 1.4b).

The level-2 cleaned events files are used in the {\tt XSELECT} tool distributed with {\it heasoft} to extract source light curves, images and spectra. 
The clean process removed events during the occultation by the Earth, any contamination by the charged particles due to passage of the satellite through the South Atlantic Anomaly region and selected events with grade 0–12 (single-quadruple events). The filter ``pha\_cutoff" is applied to select various energy channels [e.g., channel 30-700 for 0.3-7.0 keV band] for the light curves. 

The pile-up effect is very common for the CCD based detectors, however, due to the large PSF of the SXT, it becomes effective only for extremely bright sources like Crab or brighter (source count rate $>$ 180 s$^{-1}$). \source\ being only a moderately bright (max SXT count rate $\sim$ 25 s$^{-1}$, Ref. Figure \ref{model2017}) has no detectable pileup issue. A circular region of 16 arcmin radius around the source location, which encircles more than 95\% of all photons as estimated by the standard {\tt sxtEEFmake} module distributed through the POC website, is used to extract the source spectra and light curves. The appropriate ARF file suitable for the specific source region is generated using the command line auxiliary tool {\tt sxtARFModule}. Because of the large point spread function (PSF) of SXT, we are unable to extract background products from the same frame and hence the background spectrum provided by the POC\footnote{https://www.tifr.res.in/~astrosat\_sxt/dataanalysis.html} is used for spectral modelling. The background correction in the lightcurves corresponding to various energy bands is done by subtracting a constant which is the rate of background counts for a specific energy band normalized to the area of the source region, estimated by importing the background spectrum in {\tt XSPEC} (version 12.10.1) and applying the corresponding energy filter. This background  spectrum is extracted using the data from various blank-sky observations at various locations in the sky taken over the first 3 years of \astrosat\ operations. 
The light curves thus generated are re-binned, and hardness ratio plots are generated using the {\tt lcurve} utility of the {\it heasoft} package. The spectral modeling of the SXT spectra is performed using {\tt XSPEC}. The nH column density is fixed to 1.07$\times$10$^{21}$ cm$^{-2}$ as estimated by the web-based tool of the Leiden/Argentine/Bonn (LAB) Survey of Galactic HI in the direction of \source\ (beam size of 0.266$^\circ$) throughout this work to account for the Galactic photoelectric absorption in Milky Way. Temporally resolved SXT spectra are extracted and modelled to study the temporal variability of the spectral parameters (see \S \ref{subsec:timeresspec} for more details). 

\subsubsection{{\it AstroSat}-Large Area Proportional Counters (LAXPC)}    

\astrosat\ hosts three identical units of proportional counters, filled with highly pressurized Xenon gas, in a specific arrangement to provide collective effective area of 6000 cm$^2$. This instrument is non-focusing and has a field of view of $\sim$1$^\circ$ $\times$ $\sim$1$^\circ$. It is sensitive mainly in the 3.0$-$80.0 keV band \citep[][]{2016SPIE.9905E..1DY, 2017ApJS..231...10A}. 

The LAXPC field of view axis is nearly coincident to the other on-axis instruments on board \astrosat, namely CZTI, SXT and UVIT. Thus all sources observed with the SXT as the prime instrument are automatically observed by the hard X-ray detector LAXPC. The {\astrosat}-LAXPC observations of \source\ performed at various epochs are analysed using the recent LAXPC pipeline package {\tt laxpcSoft} managed and distributed by the LAXPC POC\footnote{https://www.tifr.res.in/~astrosat\_laxpc/software.html} in Mumbai, India. The background models, response functions, and gain variations are appropriately applied to generate the multi-band light curves and spectra. The modelled background spectrum is properly shifted for the gain values appropriate for the time of observations using the command line utility {\tt gainshift}, distributed with {\tt laxpcSoft}. The resulting spectra and light curves are then used for further investigations utilizing {\tt XSPEC} and {\tt lcurve}. The details of observations are listed in Table.\ref{tab:timesegref}.     

\subsubsection{{\it Neil Gehrels Swift Observatory} X-ray Telescope (XRT)\label{subsec:swiftobs}}
\swift\ has performed a number of observations covering the duration of interest to this work, sometimes even overlapping with the \astrosat\ pointings (See Table. \ref{tab:swtspec_par} for details of Swift observations analysed). The X-ray data from XRT were reprocessed with the mission-specific {\it heasoft} tool {\tt xrtpipeline} (version 0.13.4) with standard input parameters as recommended by the instrument team. This step generates new cleaned events files with the most recent calibrations. The events files thus generated are used in the multi-mission tool {\tt XSELECT} for extracting source and background products. We have analysed the data taken in both operational modes, namely the PC and WT modes. For PC mode data an annular region centred at ($\alpha$=19:59:59.929, $\delta$=+65:08:54.65), with an inner radius of 10$^"$ and outer radius of 70$^"$ is used as source region. Whereas, for the background,  another annular region centred at same location but with inner and outer radii of 150$^"$ and 350$^"$, is used. The choice of an annular source region for the PC mode data is made to mitigate the pile-up effect because for all the observations in this mode show count rates $>$ 0.6 c/s. We cross-checked for the presence of another X-ray source contaminating the source or background regions. The choice of the source region for the WT mode observations is made following the recommendations by the instrument team\footnote{https://www.swift.ac.uk/analysis/xrt/index.php}. A circular region of radius 64$^"$ centred at the location of \source\ is used for extracting source products. An annular region centred at \source\ with inner and outer radii of 188.59$^"$ and 282.88$^"$, respectively, is used for the extraction of the background. This background region selection ensures symmetrical placing about 100 pixels (the half-width of the WT window) and hence no matter where the source is in the WT window, the background region will contain r$_2$ - r$_1$ - 1 (where r$_1$ \& r$_2$ are the inner and outer radii of the annular region) pixels in 1D (minus one, as the end-of-window pixels are flagged as bad by the ground software processing and are therefore not available for use\footnote{https://www.swift.ac.uk/analysis/xrt/backscal.php}). 
The spectra and multi-band lightcuves thus generated are used in {\tt XSPEC} and {\tt lcurve} for high end investigations.  It should be noted that the BACKSCAL keyword of the source and background spectra are edited to proper values, as applicable to the current source and background source selections, before performing the spectral analysis.

\subsubsection{Short-term time-resolved spectra from XRT + SXT} \label{subsec:timeresspec} 

Around 400 \swift-XRT spectra taken between 2015.0$-$2021.2 are analyzed using the absorbed Log Parabola photon spectrum model to obtain the spectral changes during various flux states of the source. Additionally, the \astrosat-SXT observations are split into segments of $\sim$3500s duration in order to generate time-resolved spectra, which are then fitted with the aforementioned model (see \S\ref{subsec:xrayspecfit} for the model). The time-resolved spectra from SXT are extracted by applying time filtering in {\tt XSELECT} using different merged cleaned events files (one merged file for each individual observation). The total SXT observations thus yield 81 spectra with exposure times $\geq$1500s. A similar splitting of the LAXPC observations in such small time bins results in poor spectral data and hence cannot constrain the spectral shapes beyond 8.0 keV. Therefore, the LAXPC spectra are not used for this part of the study. The best-fit XRT model parameters are obtained in the 0.3-10.0 keV band, whereas the unabsorbed fluxes are estimated for the common energy band i.e., 0.3-7.0 keV, in order to combine the flux estimations from the two X-ray instruments. Tables \ref{tab:astspec_par} and \ref{tab:swtspec_par}, available only as supplementary material, provide the details of the best fit parameters.
     
\subsection{{Optical/UV Observations}}

\subsubsection{{\it Neil Gehrels Swift Observatory}-UVOT}

The UVOT observations in six optical/UV filters for all the relevant observations listed in Table \ref{tab:swtspec_par} are analysed using recent mission specific tools such as {\tt uvotimsum}, {\tt uvotsource} and {\tt uvot2pha} distributed with the {\it heasoft} package. The sky images in a particular filter corresponding to individual observations are combined using {\tt uvotimsum} to get a single frame per observation, whenever more than one image was taken. The combined images are then analysed utilizing the tool {\tt uvotsource} using a circular region of 5" radius centred at the sky location of \source\ as source region. Another circular region of 35.76" located in a source free region around 3.5' away from \source\ is used to extract background counts. 

A correction due to reddening, E(B-V)=0.178, due to the presence of the neutral hydrogen along the line of sight within our own Milky Way Galaxy, is applied to the fluxes before using these values into SEDs. The reddening is estimated by the Python module {\tt extinctions} using the two-dimensional dust map of the entire sky by \citet{1998ApJ...500..525S} which was recently updated by \cite{2011ApJ...737..103S} [SFD hereafter]. The estimation of the same parameter using the two-dimensional dust map at NASA/IPAC archive\footnote{https://irsa.ipac.caltech.edu/frontpage/} yields a value of 0.172. We also estimate this parameter using the recent three dimensional dust map by \cite{2015ApJ...810...25G} which turns out to be 0.180. This implies that we can safely use 0.178 measured using SFD. The empirical formalism by \cite{1989ApJ...345..245C} 
with A$_V$ = R$_V$ * E(B-V) and R$_V$ = 3.1 is used to estimate the correction factor A$_\lambda$ for individual UVOT filters.  
Multiple UVOT observations taken during periods over which individual SEDs were collected, were averaged to give one data point per filter.  

\subsubsection{ Mt. Abu Infrared Observatory, India}

In addition to the optical/UV observations from the {\it Neil Gehrels Swift Observatory}-UVOT, optical photometry observations from the Mt. Abu Infrared Observatory (MIRO) are used in this investigation. 

A number of optical photometric observations were made during several epochs between December 2015 and December 2017 using the \miro, including several observations contemporaneous to the \astrosat\ monitoring in 2016. Table \ref{tab:obslog} provides further information about these observations. The data were obtained using the EMCCD based optical camera installed at the f/13 cassegrain focus of the 1.2 m telescope. The data reduction and the photometry procedures adopted are discussed in \citet[][]{Kaur2017}. 
  Differential photometry, using several comparison stars in the same frame as the source, was used to minimize atmospheric seeing effects. The calibrated magnitudes thus obtained were converted to the fluxes and corrected for Galactic extinction. The nightly averaged fluxes in mJy are shown in the bottom panel of the Figure \ref{fig:mwllc}.
       
\subsection{Other Publicly available Resources}
 
For coverage at lower frequencies, we use the publicly available radio data from the Owens Valley Radio Observatory (OVRO)\footnote{https://sites.astro.caltech.edu/ovroblazars/index.php?page=home} at 15 GHz\citep{2011ApJS..194...29R}. For completeness of the SED, we extract the quasi-simultaneous TeV data from \cite{MAGIC2020}, obtained by the MAGIC telescope, which have been corrected for $\gamma-\gamma$ absorption by extra-galactic background light (EBL). The TeV $\gamma-$ray quick look lightcurve from the FACT \citep[First G-APD Cherenkov Telescope;][]{2013JInst...8P6008A, 2014JInst...9P0012B, 2015arXiv150202582D}  during September 2016 to November 2017 are also used to investigate the high energy counterparts of the observed X-ray activities.  

 \label{subsec:obs}

\section{Results} \label{sec:Results}

 This section presents the results of our comprehensive multi-wavelength study of flux and spectral-variability of \source\ using \astrosat\ and other facilities. 

\begin{figure}
\includegraphics[scale=0.32]{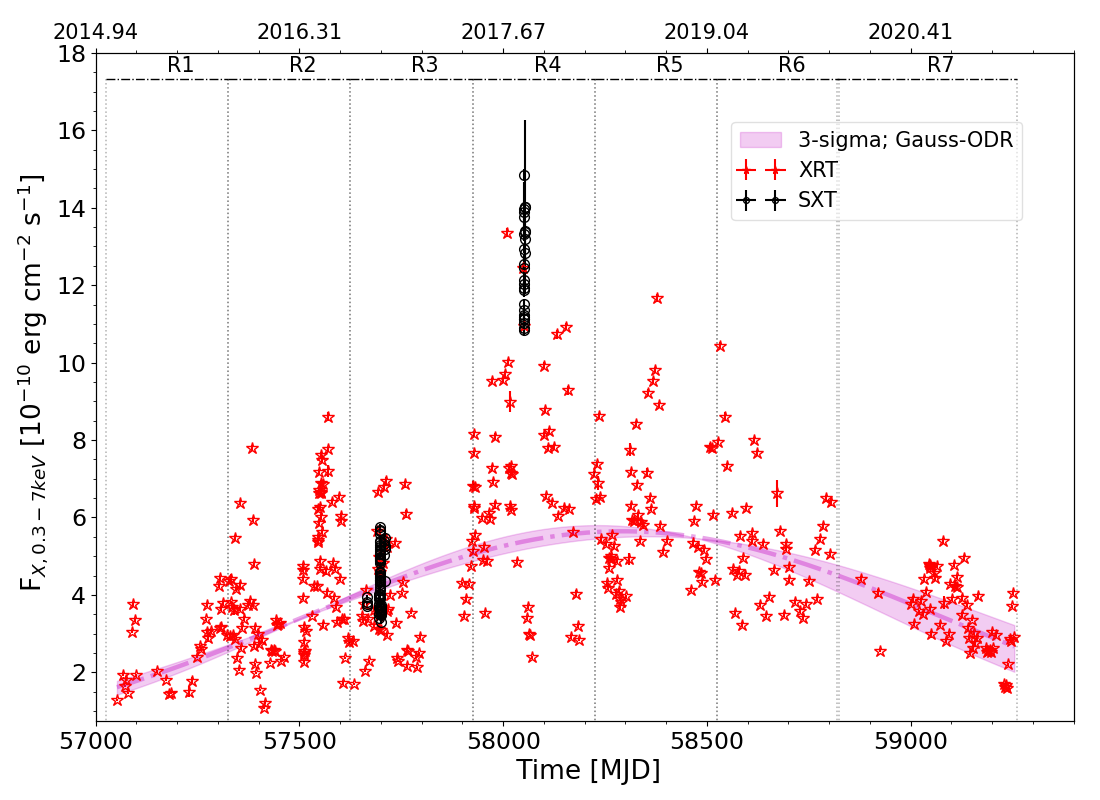}
\caption{\label{fig:longlc}{\footnotesize X-ray light curves in the energy range of 0.3-7.0 keV obtained using XRT and SXT data from 2015 December to February 2021. The red stars represent XRT fluxes, whereas the black open circles show SXT fluxes. The shaded region around the best fit dotted curve represents the 3$\sigma$ confidence interval. }}
\end{figure}

\subsection{Light Curves and Flux Variations}
\label{sec:lcflxvar}
X-ray light curve in the energy band of 0.3-7.0 keV taken over ~6 years (between MJD 57000 to 59260) is shown in Figure \ref{fig:longlc}. The open stars symbolize the integrated fluxes from XRT, whereas the open circles represent the fluxes from SXT. The long SXT exposures are split into several small time intervals of 3500s, within which spectra are extracted and the best fit fluxes are used to construct the light curve (see \S \ref{subsec:timeresspec} for details). However, in order to generate the light curve from XRT, the fluxes are extracted from the best fit model spectra from the individual observations between January 2015 to February 2021. 

The overall long-term average flux variation trend which is mathematically characterized by a broad Gaussian peaking at $\sim$ MJD 58233 and FWHM of $\approx$ 785 days. The long-term trend is superimposed by several flares.

The 6 years long XRT lightcurve (Fig.\ref{fig:longlc}) is divided into seven segments of 300 days each except the last one which corresponds to 440 days. These segments, R1, R2, R3, R4, R5, R6 and R7, encompass a number of X-ray flares at different epochs sampling the different parts of the above said Gaussian function [See Fig. \ref{fig:longlc}].  The \astrosat\ observations from 2016 and 2017 fall into R3 and R4, respectively.  The following paragraphs summarize the quantitative analysis of the multi-wavelength variability during various X-ray flares around the \astrosat\ monitoring.  

\begin{figure*}
    \centering
    \includegraphics[trim={0.2cm 1.85cm 1cm 1cm}, clip=true,width=0.495\textwidth]{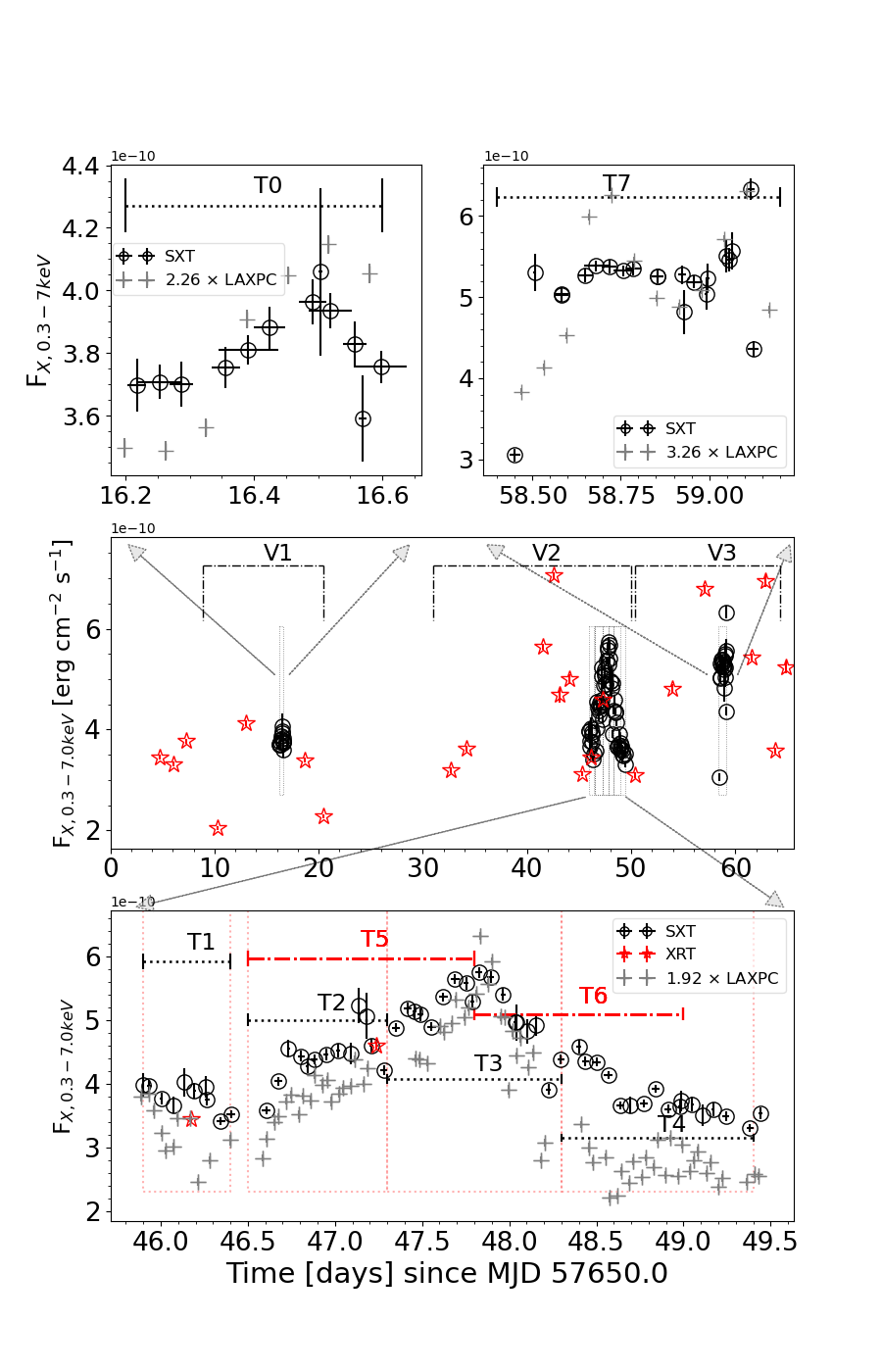}
    \includegraphics[trim={0.2cm 0.65cm 1cm 1cm}, clip=true,width=0.495\textwidth]{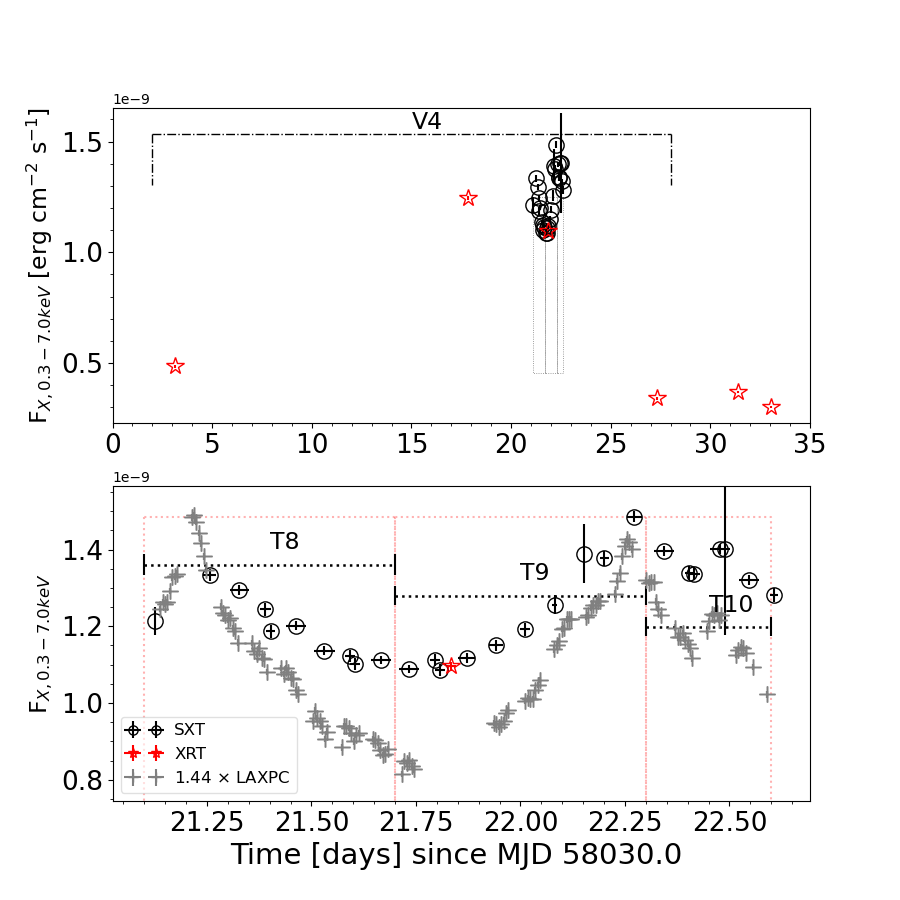}
    \caption{Zoomed-in light curves focusing on the time around the flares observed by \astrosat\ in 2016 and 2017. {\bf Left, hereafter Fig.\ref{fig:zmdlc4long}a}: Various time segments (V1,\ldots,V3) and subgroups (T0,\ldots,T7) during 2016, zooming into the flaring activities in {\bf the top and bottom panels}. {\bf Right, hereafter Fig.\ref{fig:zmdlc4long}b}: Time segment V4 and various subgroups (T8,T9,T10) during 2017, zooming into the bright flaring activities in the bottom panel. The LAXPC light curves are also shown by grey '+' markers in the zoomed windows after scaling up to fit in the panels.}  
    \label{fig:zmdlc4long}
\end{figure*}

The SXT light curves reveal that \source\ exhibits significant flux variations with different time-scales at all the epochs as shown in Figure \ref{fig:zmdlc4long}. Therefore, in order to understand the nature of the flux variability during and around the \astrosat\ observations, four time-segments are created. The basic criteria behind the division of these variability profiles is to distinguish and characterize the X-ray outbursts (doubling/halving time scales \& peak flux) which are probably related to the same physical processes which triggered the X-ray activities recorded by the \astrosat. These segments are termed as variability profiles and are denoted by V1, V2, V3 and V4 [See Fig. \ref{fig:zmdlc4long}]. Note that the V1, V2 and V3 are subsets of R3, while V4 is a subset of R4. In order to investigate the time dependent spectral behaviour of the flaring activities observed with \astrosat\, small portions of V1, V2, V3 and V4 are further subdivided into time-segments denoted by T0, T1,\ldots,T10 [See Fig.\ref{fig:zmdlc4long} and Table\ref{tab:timesegref} for the details]. These segments zoom-in on to the flux variations during the \astrosat\ monitoring. Another \astrosat\ monitoring data with total exposure of $\sim$ 2.9 ks from November 2015, denoted by `PV', is also included in this list to compare the X-ray activities at earlier epochs [See \S\ref{subsec:xrayspecfit} for details]. 

\begin{table*}[t]
\caption{Estimation of $\Delta t$ and $F_{var}$}
{\footnotesize \begin{tabular}{llllll}
\hline\hline
{\bf T$_{tag, prof.}$} & {\bf Duration} &  {\bf $\Delta$t$_D$/$\Delta$t$_H$ }  &  {\bf F$_{var, com}$/\bf F$_{var, SXT}$}  & F$_{X,p}$& {\bf T$_{tag, SXT}$}  \\
&    [MJD]   &  [Days]        &  & [erg cm$^{-2}$ s$^{-1}$]         &  \\
                   &      &         &  & $\times 10^{-10}$         &  \\
     \hline
{\bf V1} & {\bf 57660.0 - 57670.0} & $\uparrow_{XRT}$: 2.69$\pm$0.02 & $_{COM.}$: 0.25$\pm$0.005  & 4.12 &  \\
   & & $\downarrow_{COM.}$: 8.68$\pm$0.05 &   & & \\
   & 57666.22-57666.63& $\uparrow_{SXT}$: 3.09$\pm$0.17 & $_{SXT}$: 0.02 $\pm$0.004  &  & T0 \\
   & & $\downarrow_{SXT}$: 1.69$\pm$0.01 &   & &  \\
   &  & $\uparrow_{LAXPC}$: 1.01$\pm$0.008 & $_{LAXPC}$: 0.07$\pm$0.002  &  &  T0\\
   & & $\downarrow_{LAXPC}$: 0.63$\pm$0.01 &   & &  \\  
\cmidrule{2-6} 
{\bf V2} & {\bf 57682.6 - 57700.0} & -- &  $_{COM.}$: 0.19$\pm$0.002 &  & \\
    SF1        &   57682.70–57695.25   & $\uparrow_{XRT}$: 8.59$\pm$0.06 & $_{SF1}$: 0.31$\pm$0.006 & 7.05 &\\
            &                   & $\downarrow_{XRT}$: 2.31$\pm$0.01 & &  &\\            
SF2   &   57696.30-57700.00     & $\uparrow_{COM.}$: 2.92$\pm$0.003 & $_{SF2}$: 0.15$\pm$0.003 & 5.74 & T1+T2+(T3/2) \\
                &                         & $\downarrow_{COM.}$: 1.95$\pm$0.01  & & &     (T3/2)+T4     \\ 
                                    & 57696.46-57700.00 & $\uparrow_{LAXPC}$: 1.49$\pm$0.01 & $_{LAXPC}$: 0.26$\pm$0.001  &  & T1+T2+(T3/2) \\
                                    & & $\downarrow_{LAXPC}$: 0.65$\pm$0.009 &   & &  (T3/2)+T4 \\  
\cmidrule{2-6}          
{\bf V3}          & {\bf 57700.36 - 57714.3} &  -- &  $_{COM.}$: 0.18$\pm$0.004  & &  \\
      SF1      &        57700.36–57709.40          & $\uparrow_{XRT}$: 5.93$\pm$0.07   & $_{SF1}$: 0.30$\pm$0.007 & 6.78 &\\
            &                  & $\downarrow_{COM.}$: 5.19$\pm$0.08  &   & & \\ 
         SF2   &           57709.40–57713.80       & $\uparrow_{COM.}$: 8.89$\pm$0.03 & $_{SF2}$ 0.26$\pm$0.009& 6.94 &\\
            &                  & $\downarrow_{XRT}$: 1.10$\pm$0.001 & & &\\  
            &    57708.45-57709.11   & $\uparrow_{SXT}$: 1.23$\pm$0.001  & $_{SXT}$: 0.09$\pm$0.003   & & T7\\  
             &                       &                   --                & $_{LAXPC}$: 0.15$\pm$0.002   & & T7\\
          &  & $\uparrow_{LAXPC}$: 0.36$\pm$0.003 & $_{LAXPC,SF1}$: 0.17$\pm$0.003  &  &  \\
                                    & & $\downarrow_{LAXPC}$: 0.53$\pm$0.005 &   & &  \\
          & & $\uparrow_{LAXPC}$: 0.52$\pm$0.004 & $_{LAXPC,SF2}$: 0.12$\pm$0.003  &  &  \\
                                    & & $\downarrow_{LAXPC}$: 0.18$\pm$0.003 &   & &  \\\hline
{\bf V4}          & {\bf 58032.0 - 58058.0} &                                  -- & $_{COM.}$ : 0.18$\pm$0.002 & & \\
 SF1           &                 & $\uparrow_{XRT}$: 10.81$\pm$0.04 & $_{SF1}$: 0.56$\pm$0.005 & 12.45 & \\
            &                  & $\downarrow_{XRT}$: 5.06$\pm$0.03 &  &  &  \\
           & 58051.1-58052.6  & -- & $_{SXT}$: 0.09$\pm$0.008  &  & T8+T9+T10  \\
            & 58051.1 - 58051.7        & $\downarrow_{SXT}$: 2.06$\pm$0.03 & $_{T8}$: 0.07$\pm$0.003 &  & T8 \\            
   SF2         & 58051.7 - 58052.3                  & $\uparrow_{SXT}$: 1.14$\pm$0.01 & $_{T9}$: 0.10$\pm$0.009 & 14.8 & T9 \\
            &  58052.3 - 58052.6                 & $\downarrow_{SXT}$: 1.60$\pm$0.009 & $_{T10}$: 0.04$\pm$0.004 &  & T10 \\
            &                      & -- & $_{LAXPC}$: 0.15$\pm$0.003  &  & T8+T9+T10  \\
                                &  & $\downarrow_{LAXPC}$: 0.63$\pm$0.004 & $_{T8}$: 0.17$\pm$0.001  & &  T8\\
          &  & $\uparrow_{LAXPC}$: 0.54$\pm$0.003 & $_{T9}$: 0.13$\pm$0.001 &  &  T9\\
                                    & & $\downarrow_{LAXPC}$: 0.46$\pm$0.003 & $_{T10}$: 0.07$\pm$0.001  & &  T10\\
\hline 
\end{tabular}} \\
\label{tab:tvar}
Note: T$_{tag, prof.}$ and T$_{tag,SXT}$ refer to the tags adopted for defining the variability profiles and time-segments within the SXT light curves, respectively [Ref. Fig.\ref{fig:zmdlc4long}a, \ref{fig:zmdlc4long}b for the details]. SF represents a small flare observed in each segment, where $\uparrow$ and $\downarrow$ represent its doubling ($\Delta t_D$) and halving ($\Delta t_H$) timescales, estimated for various flares. The subscripts XRT, SXT, LAXPC and COM., refer to the data used from XRT, SXT, LAXPC and combined from the both XRT/SXT, respectively. The F$_{X,p}$ represents the peak flux in a particular time segment. 
\end{table*}

The characteristic flux doubling/halving timescales ($\Delta t_D$/$\Delta t_H$) of the flares in the different variability profiles (V1, V2, V3, V4), using the combined X-ray lightcurve from both SXT \& XRT, are derived with $\Delta t = t_d\,\times\,ln\,2/|ln(F2/F1)|$ \citep{2013ApJ...766L..11S}. Here, $F1$ and $F2$ are the fluxes observed at a time interval of $t_d$. The methodology is also applied to the detailed X-ray lightcurve from SXT to obtain the parameters over shorter time-scales [See Table \ref{tab:tvar} for details]. The variations are further characterised by the fractional variability amplitude ($F_{var}$) defined in \citet{2003MNRAS.345.1271V} and given by

\begin{equation}
F_{var} = \sqrt{\frac{S^2 - \sigma^2_{err}}{x_m^2}} 
\end{equation}
$\sigma^2_{err}$ is the mean square error of each observations and $S^2$ is the sample variance, where $\sigma_{XS} = S^2 - \sigma^2_{err}$ is the excess variance. $x_m$ is the unweighted sample mean for N points. The error in $F_{var}$ is given by
\begin{equation}
\sigma_{F_{var}} = \sqrt{ \left(\sqrt{\frac{1}{2N}}\,\frac{\sigma^2_{err}}{x_m^2 F_{var}}\right)^2 + \left(\sqrt{\frac{\sigma^2_{err}}{N}}\,\frac{1}{x_m}\right)^2} 
\end{equation}

The variability time scales ($\Delta$t$_D$/$\Delta$t$_H$), $F_{var}$, and peak flux (F$_{X,p}$) are derived for the flares in each time-segment (V1, V2, V3, V4) of the long light curves shown in Fig. \ref{fig:zmdlc4long}, including both XRT and combined XRT-SXT observations. The values are reported in Table \ref{tab:tvar}.  

\begin{figure*}
\includegraphics[trim={0cm 1cm 2cm 2cm}, clip=true, width=0.5\textwidth]{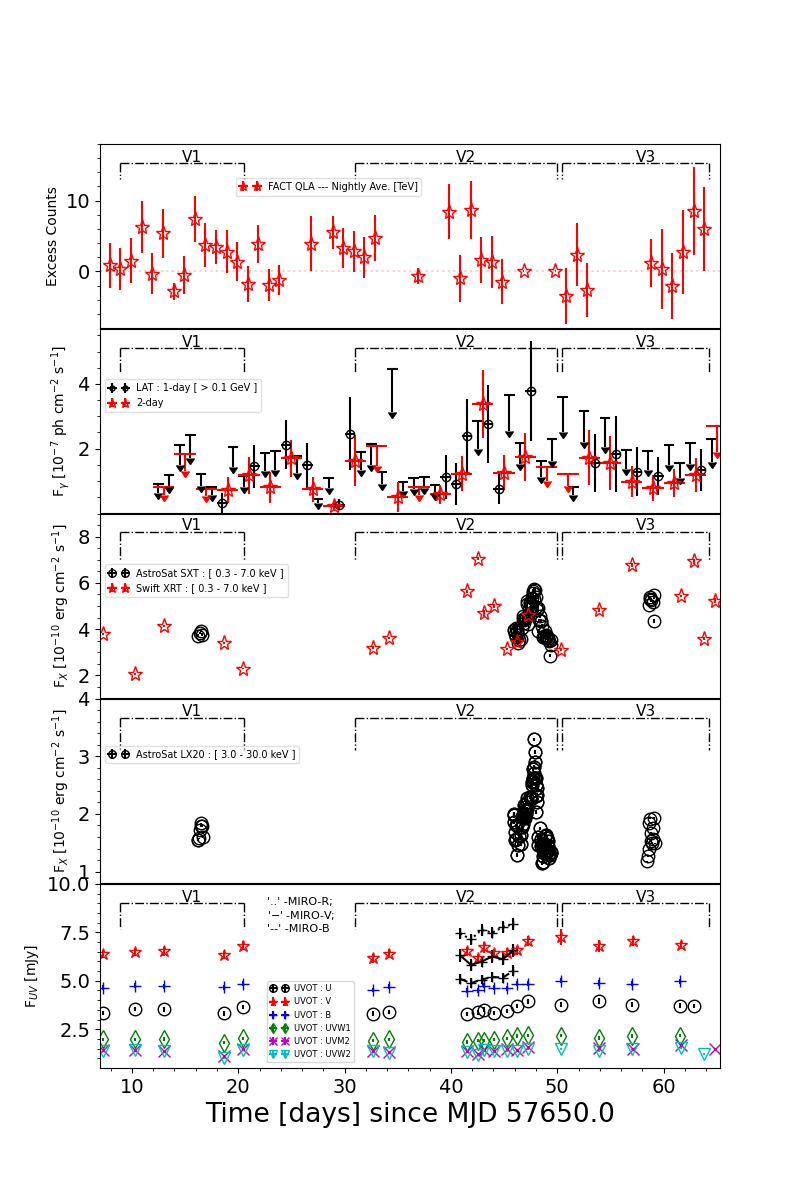}
\includegraphics[trim={0cm 1cm 2cm 2cm}, clip=true, width=0.5\textwidth]{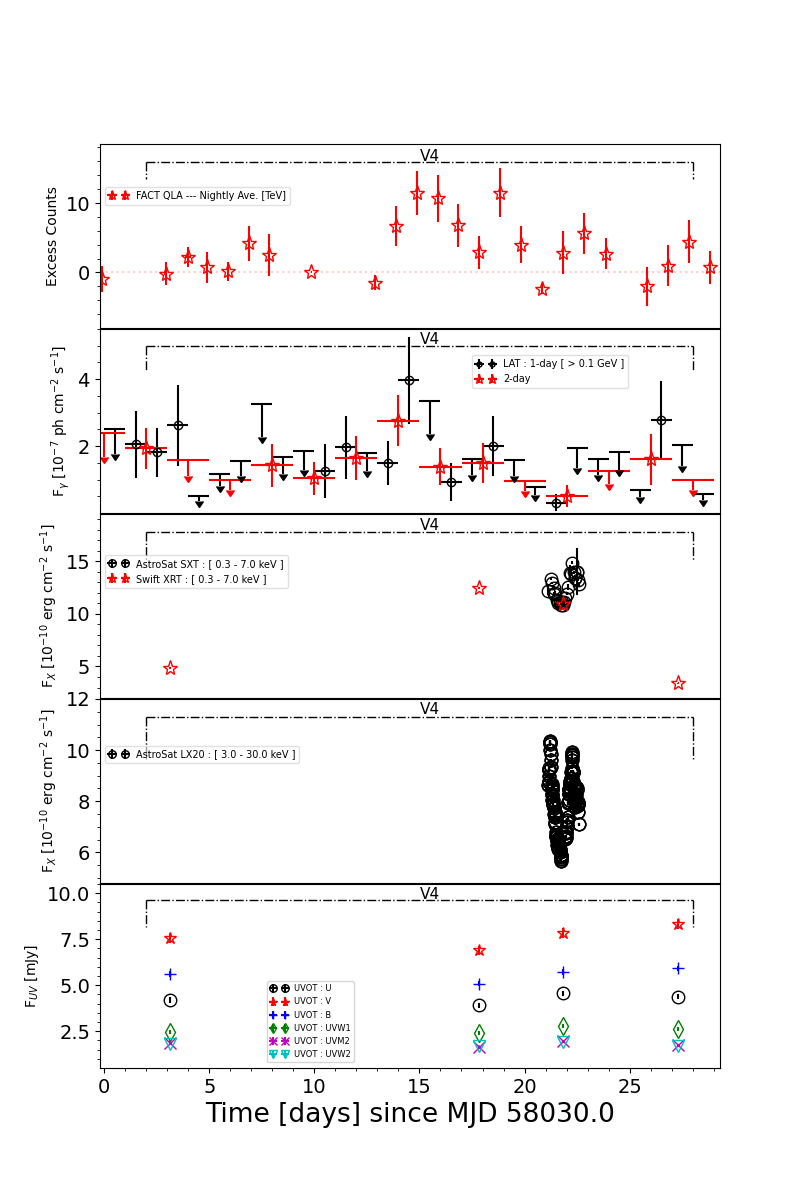}

\caption{\label{fig:mwllc}{\bf Left: hereafter Fig.\ref{fig:mwllc}a} Multi-wavelength light curves of {\source}. From top to bottom:  {\bf FACT} (quick look analysis, binned nightly), {\bf \emph{Fermi}-LAT} (binned in 1 day and 2 days), {\bf SXT} and {\bf XRT} (binned by orbit), {\bf LAXPC} and {\bf UVOT} (U, V, B, UVW1, UVM2 and UVW2 bands) and {\bf MIRO} (B,V and R bands, binned nightly) during October-November 2016 (corresponds to segments V2 and V3). {\bf Right: hereafter Fig.\ref{fig:mwllc}b} Same as left, but for the observations during November 2017 (corresponds to segment V4). Note that 95~\% upper limits instead of flux points are used for time-bins where the test statistics (TS) in the LAT light curves are TS$\le$25.} 
\end{figure*}

The multi-wavelength light curves shown in Fig. \ref{fig:mwllc} illustrate the prominent X-ray activities and its counterparts in other energy bands. Some of the X-ray outbursts seem to have associated high energy counterparts in GeV and TeV bands. There have been several communications from the FACT collaboration through `Astronomers Telegrams (ATels)' \citep{2016ATel.9010....1B, 2016ATel.9203....1B, 2016ATel.9148....1B,2016ATel.9139....1B, 2016ATel.9239....1B} reporting fluxes beyond 1 Crab unit (CU), and 36 private communications to the partners to trigger multi-wavelength observations during moderately high flux states (F$_\gamma$ $\ge$ 0.5 CU) over the period of 2016-2017. The FACT quick look analysis (QLA) light curve\footnote{https://fact-project.org/monitoring} of \source\ shows several flaring episodes.
In the following subsections, the detailed variability profiles, the nature of the X-ray flux variations and its multi-wavelength associations are discussed.

\subsubsection{Variability profile 1 (V1)}

The variability profile V1 refers to the light curves around T0 starting from MJD 57660.0 to 57670.0. It also includes four pointings by \swift. The X-ray light curves (see Fig. \ref{fig:zmdlc4long}a) show that the observations performed during T0 have been part of a fast varying flux state (F$_{var}$ $\sim$ 0.25$\pm$0.005) with initial rise ($\Delta$t$_D$ $\sim$ 2.7 days) and fall ($\Delta$t$_D$ $\sim$ 8.7 days). The X-ray light curves in 0.3-7.0 keV (SXT) and 3.0-30.0 keV (LAXPC) bands over T0 show significant flare-like variations (See Table \ref{tab:tvar} for details). The flux variation in the 3-30.0 keV band is more prominent than in the 0.3-7.0 keV band. As shown in Fig\ref{fig:mwllc}a, V1 comprises hints of flux variations in the TeV band whereas no variations are observed in the optical/UV and GeV bands.  

\subsubsection{Variability profile 2 (V2) }

V2 corresponds to the light curves during MJD 57682.6 and 57700.0 (see Fig. \ref{fig:zmdlc4long}a) which starts 12.5 days after the end of V1. The combined light curves clearly show that V2 encompasses two consecutive X-ray flares with the first one barely covered by XRT (SF1) and the second one entirely observed by the \astrosat\ (SF2). The peak to peak time difference of the SF1 and SF2 is $\sim$ 6.9 days. 

SF1 is highly asymmetric in X-rays, and shows noticeable TeV and GeV activities. SF2, however, while also highly asymmetric in X-rays, hardly shows any variations at other energies. The fastest variations during V2 are observed with LAXPC during SF2 (${\Delta}t_D$ = 1.5 days and ${\Delta}t_H$=0.65 days). Hence, the X-ray variations during both SF1 and SF2 can be characterized by `slow rise and fast decay' profiles [See Table \ref{tab:tvar}].
 
The \astrosat\ light curves are further subdivided into time-segments T1, T2, T3, T4, T5, and T6 [See Fig. \ref{fig:zmdlc4long}a]. In addition to the average flux variations over the \astrosat\ observations, the variability time scales and fractional variability indicate significant variations even during above mentioned T-segments [See Table \ref{tab:tvar}]. Combining the optical/UV fluxes from MIRO and UVOT, we deduce that no significant optical and UV counterparts of the X-ray flares during V2 are seen [Ref. Fig.\ref{fig:mwllc}a]. 
 
\subsubsection{Variability profile 3 (V3)}

The variability profile V3 extends from MJD 57704.3 - 57714.3, that is, it starts soon after the end of V2. The combined XRT and SXT X-ray light curves during V3 (see Fig. \ref{fig:zmdlc4long}a), show that \astrosat\ pointing T7 is most probably the falling part of an X-ray flare which peaks at the flux (6.78 $\times$ 10$^{-10}$ erg cm$^{-2}$ s$^{-1}$) comparable to the same of V2-SF1. The V3 profile contains two (SF1 and SF2) X-ray flares of nearly similar peak fluxes with the second one decreasing very fast (${\Delta}$t$_H$ = 1.1 day). Both SF1 and SF2 show the `slow rise and fast decrease' already seen in V2. A similar trend is also seen in the LAXPC light curve (3.0-30 keV band), however, the fractional variability is higher than in the SXT band [See Table \ref{tab:tvar}]. There are no obvious counterparts in other energy bands, with the noteworthy exception of a TeV enhancement during SF2.

\subsubsection{Variability profile 4 (V4)}

The profile V4 extends from MJD 58032.0 to 58058.0 (see Fig. \ref{fig:zmdlc4long}b). This time segment signifies the X-ray flux variability around the major activity observed in November 2017. 

V4 includes 4 pointings of \swift\ with one coinciding with the \astrosat\ observations. The combined X-ray light curve during V4 shows that the X-ray flare observed with \astrosat\ is part of a prominent MWL activity lasting over $\sim$ 24 days. The \swift\ light curve alone shows that the source has doubled its flux in merely 15 days from MJD 58033.0 to 58048.0, reaching a record flux of 12.45 $\times$ 10$^{-10}$ erg cm$^{-2}$ s$^{-1}$ in the 0.3-7.0 keV band [Ref. Fig. \ref{fig:zmdlc4long}b]. The \astrosat\ pointing was made 3 days after the XRT peak, that is MJD 58051.0. The \astrosat\ observations in V4 reveal the highest flux state ever observed (14.8 $\times$ 10$^{-10}$ erg cm$^{-2}$ s$^{-1}$).
 
The observations with \astrosat\ lasting $\sim$ 2 days reveal two major sub-flares. The doubling and halving times indicate a slow decline and fast rise in them. Figure \ref{fig:peakshiftflux} and Table \ref{tab:swtspec_par} reveal a significant shift in the synchrotron peak position (E$_{s,p}$) over the flare compared to the observations in 2016.    

The LAXPC light curve shows a similar behavior as the SXT light curve but exhibits a more pronounced amplitude variability, which is underlined by a higher excess fractional variance ($EV=0.15$) compared to the SXT light curve ($EV=0.08$). As shown in Fig\ref{fig:mwllc}b, V4 comprises GeV and TeV flares nearly three days prior to the observed X-ray peak. The GeV/TeV fluxes double within $\sim$ 1 day. Interestingly, the GeV flare exhibits a narrower profile than the TeV flare and leads the TeV flare by $\sim$ 1 day. The lack of X-ray data prohibits us to judge whether the GeV/TeV peak corresponds to the (in this case unknown) X-ray maximum or if the X-ray response is delayed. The sub-flare observed with \astrosat, however, corresponds to a low state in the GeV/TeV bands. No significant UV/optical counterparts are seen. 
 
The correlated X-ray activity, along with the shift of the peak position mentioned above, implies that it's not a simple variation in electron density, but a more complicated spectral change throughout the flare that is responsible for the X-ray flare. This requires time-dependent changes in the magnetic field, the maximum electron Lorentz factor, the Doppler factor of the emission region, or a combination thereof. A more detailed analysis is provided below.

\begin{figure*}
\includegraphics[width=0.49\textwidth]{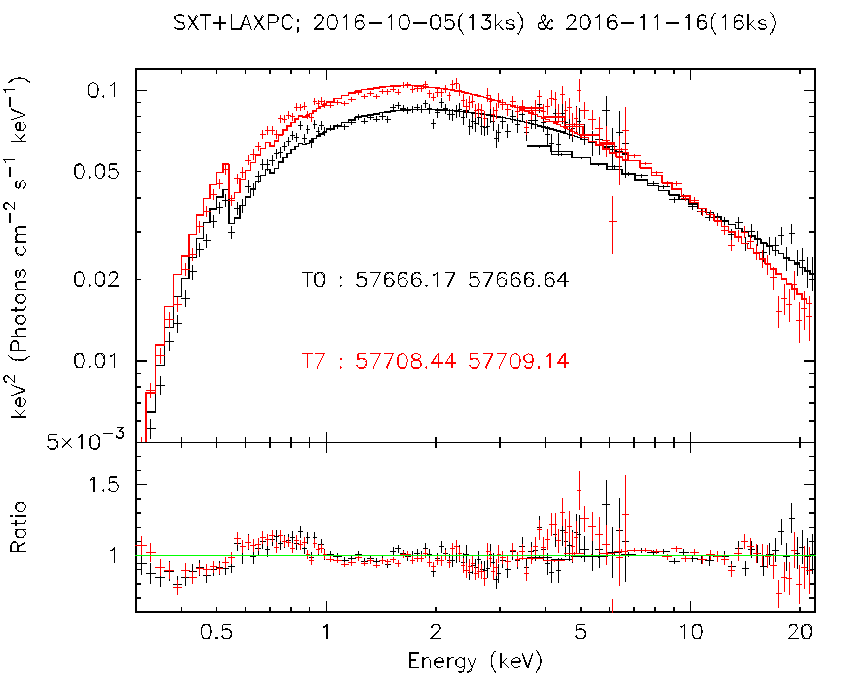}
\includegraphics[width=0.49\textwidth]{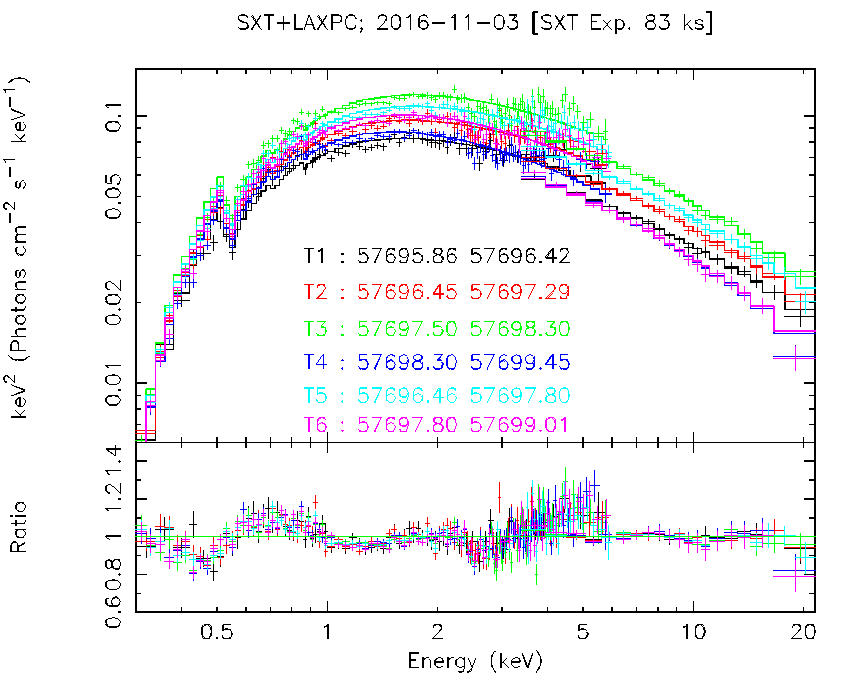}\\
\includegraphics[width=0.49\textwidth]{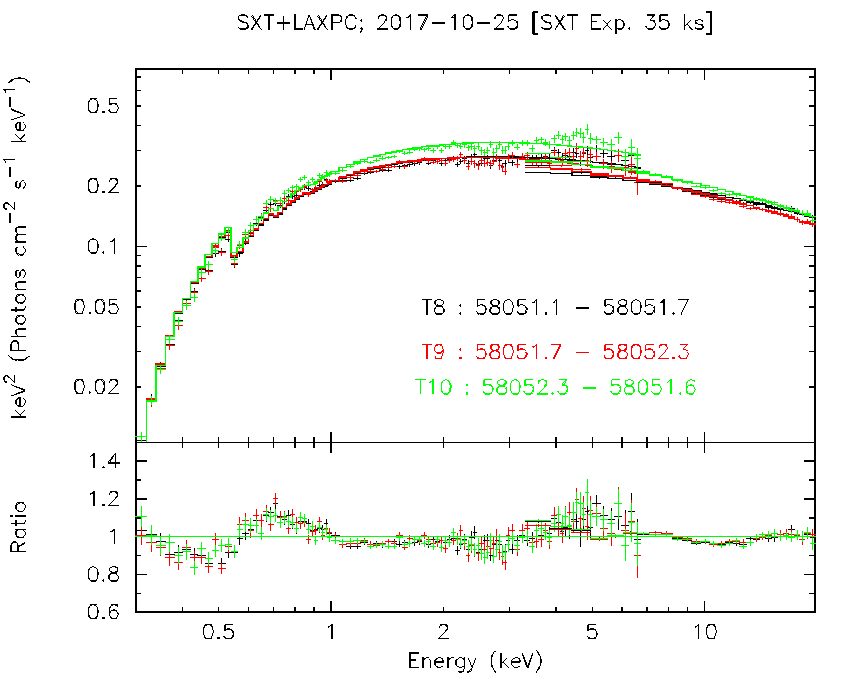}
\includegraphics[width=0.48\textwidth]{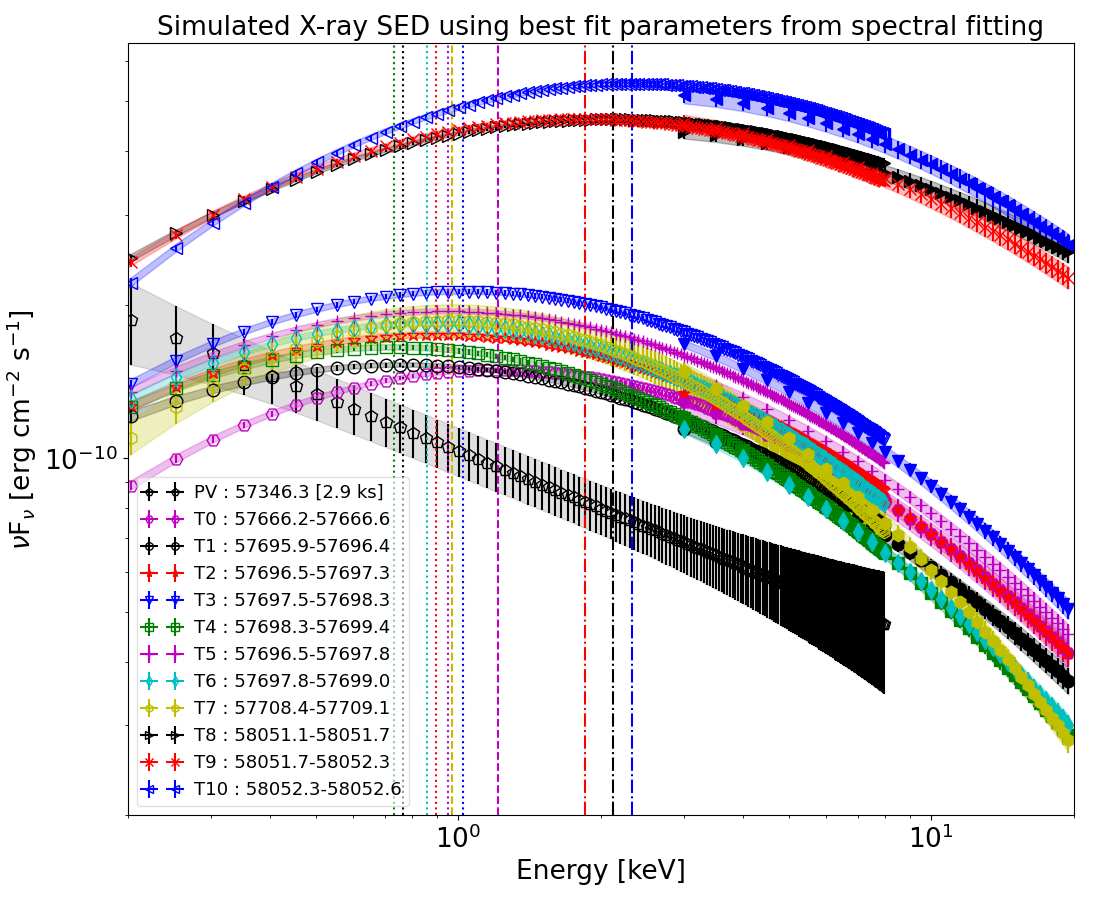}
\caption{\label{fig:specfits} X-ray spectra fitted with a log-parabola model. {\bf Top-Left, hereafter \ref{fig:specfits}a:} E$^2$F$_E$ v/s E for T0 (black) and T7 (red) extracted from \astrosat\ observations performed during 05-06 October 2016 and during 16-17 November 2016. {\bf Top-Right, hereafter \ref{fig:specfits}b:} 
Same as panel a, but for periods T1, T2, T3, T4, T5, and T6. The various spectra represent different flux and spectral states segmenting different parts of the outburst observed between MJD 57695.9 $-$ 57699.0 (03$-$07, November 2016). {\bf Bottom-Left, hereafter \ref{fig:specfits}c:}   
Same as panel a, but for T8, T9 and T10 from the outburst between MJD 58051.1 $-$58052.6, i.e. between 25-26 October 2017. {\bf Bottom-Right, hereafter \ref{fig:specfits}d:}  E$^2$F$_E$ v/s E plots with the respective butterfly diagram representing the best-fit parameters and their uncertainties for all 16 X-ray spectra shown in panels a -- d. }
\end{figure*}

\subsection{X-Ray Spectral Modeling \label{subsec:xrayspecfit}}

The combined X-ray spectra from SXT and LAXPC covering the 0.3-30 keV band from various epochs are fitted to derive the time dependent spectral behavior of the source. For these investigations, the total \astrosat\ observations are split into 12 segments (see Table \ref{tab:timesegref} and Fig. \ref{fig:zmdlc4long}). 
These time segments are designated as PV, T0, \ldots, T10. The PV segment refers to the $\sim$ 3ks \astrosat\ pointing of \source\ performed on 20 November 2015. These segments are defined to highlight the changes in the X-ray spectral parameters sampling different parts of the flares and also different average flux states [See Table \ref{tab:timesegref} and Fig. \ref{fig:zmdlc4long} for the definitions of these time-segments]. 
 
The combined SXT+LAXPC spectra from the 12 segments are individually modelled with two spectral models (1) {\tt TBabs * LogParabola}, hereafter M1, and (2) {\tt TBabs * Cuttoffpl}, hereafter M2. The absorption model component {\tt TBabs} with the WILM abundance model \citep{2000ApJ...542..914W} is used to fit the Galactic neutral hydrogen absorption in the source direction. 
The {\tt LogParabola} and {\tt CutoffPl} models fit a log-parabola and a cut-off power-law shape to the intrinsic spectra, respectively. The input parameter $nH$ is kept fixed to 0.107 $\times$ 10$^{-22}$ cm$^{-2}$ in both the models M1 and M2 throughout. This value is estimated using 21 cm observations in the source direction (GAL. Coordinate L=98.003370$^\circ$, B=17.669746$^\circ$) using the online tool\footnote{https://www.astro.uni-bonn.de/hisurvey/profile/} by the LAB Survey \citep[][and relevant reference therein]{2005A&A...440..775K} with the default 0.27$^\circ$ beam. The PV segment does not have a usable LAXPC spectrum and hence, the spectral results correspond to the SXT observations (0.3-7.0 keV) only. The $\chi^2$ statistics is used for the spectral modeling throughout. 

The best fit parameters from the spectral fitting along with their 2$\sigma$ uncertainties are listed in Table \ref{table:specfit}. The best fit results show that the observed X-ray spectra can be represented equally well by M1 and M2. We prefer M1 over M2, as (1) it has been used in previous studies, thus allowing for an easy comparison, and (2) the fitting with M2 requires higher systematic uncertainties to be added to the statistical errors to converge the fit. On the other hand, the PV spectrum is fitted with the absorbed powerlaw model ({\tt TBabs*powerlaw} with fixed nH). 

In order to get acceptable $\chi^2_\nu$ values 4-5\% systematic uncertainties are needed in many cases. Normally, 3\% systematic uncertainties are recommended for the SXT+LAXPC joint spectral fitting. Panels (a, b, c) in Fig. \ref{fig:specfits} show the observed spectra for T0 to T10 with the respective best fit models. Fig. \ref{fig:specfits}d presents the model generated spectra in the energy band 0.2$-$12.0 keV. The colored regions around the model spectra are 1$\sigma$ uncertainty intervals for the best fit parameters.

\begin{table*}
\centering
\caption{The best fit parameters for time-resolved X-ray spectra. The C parameter refers to the relative cross normalization for LAXPC keeping the SXT normalization fixed to 1.}
{\footnotesize \begin{tabular}{l|llllll|llll}
\hline
\multicolumn{1}{l}{T$_{Tag}$} &
\multicolumn{6}{c}{Log parabola } &
\multicolumn{4}{c}{Exponential Cutoff Power-law} \\
\hline
& C & $\alpha$ & $\beta$ & F$_{X, 0.3-8.0 keV}$ & E$_{s,p}$ & $\chi^2_\nu$/dof & $\Gamma$ & E$_{cut}$ & F$_{X, 0.3-8.0 keV}$ & $\chi^2_\nu$/dof \\
&  &        &         & [erg cm$^{-2}$ s$^{-1}$] &  [keV] &          &          &           &  [erg cm$^{-2}$ s$^{-1}$]  &   \\
  &   &       &         &  $\times$ 10$^{-10}$   &       &     &          &           &  $\times$ 10$^{-10}$ &   \\
\hline 
PV$^\dagger$  & - &  2.48$\pm$0.04 & -             & 2.09$\pm$0.04 & - &  1.21/197    &   2.37$\pm$0.04      &     13.68     &    2.02$\pm$0.02         &  1.31/195 \\
T0  &  0.85$\pm$0.03 & 1.98$\pm$0.02 & 0.36$\pm$0.02 & 3.17$\pm$0.02 & 1.07$\pm$0.07 & 1.6/142 & 1.97$\pm$0.02 & 13.68$^{+1.24}_{-1.06}$ & 3.26$\pm$0.02 & 1.80/142 \\

T1  & 0.85$\pm$0.03 & 2.07$\pm$0.02 & 0.33$\pm$0.02 & 3.20$\pm$0.02 & 0.78$\pm$0.06 & 1.08/311 & 2.11$\pm$0.03 & 14.75$^{+1.43}_{-1.21}$ & 3.16$\pm$0.02 &0.91/288 \\

T2  & 0.84$\pm$0.03 & 2.03$\pm$0.02 & 0.36$\pm$0.02 & 3.57$\pm$0.02 & 0.91$\pm$0.05 & 1.29/360 & 2.06$\pm$0.02 & 13.28$^{+0.99}_{-0.87}$ & 3.69$\pm$0.02 & 1.21/330 \\

T3  & 0.83$\pm$0.02 & 1.99$\pm$0.01 & 0.38$\pm$0.02 & 4.42$\pm$0.02 & 1.03$\pm$0.03 & 1.78/397 & 2.04$\pm$0.02 & 13.07$^{+0.91}_{-0.80}$ & 4.57$\pm$0.02 & 1.09/367 \\

T4  & 0.89$\pm$0.03 & 2.09$\pm$0.01 & 0.40$\pm$0.02 & 3.14$\pm$ 0.01 & 0.77$\pm$0.03 & 1.55/416 & 2.13$\pm$0.02 & 11.43$^{+0.86}_{-0.76}$ & 3.26$\pm$0.01 & 1.14/386 \\

T5  & 0.81$\pm$0.02 & 2.01$\pm$0.01 & 0.37$\pm$0.02 & 4.00$\pm$0.01 & 0.97$\pm$0.03 &1.60/456 & 2.05$\pm$0.02 & 13.10$^{+0.82}_{-0.73}$ & 4.14$\pm$0.01 & 1.27/426 \\

T6  & 0.72$\pm$0.02 & 2.05$\pm$0.01 & 0.40$\pm$0.02 & 3.64$\pm$0.01 & 0.87$\pm$0.03 & 1.67/447 & 2.08$\pm$0.02 & 11.29$^{+0.79}_{-0.71}$ & 3.77$\pm$0.01 & 1.23/412 \\

T7  & 1.03$\pm$0.04 & 1.98$\pm$0.02 & 0.52$\pm$0.03  & 3.72$\pm$0.02 & 1.05$\pm$0.05 & 1.84/132 & 1.96$\pm$0.02 & 9.41$^{+0.62}_{-0.56}$ & 3.71$\pm$0.02 & 1.70/132 \\

T8  &0.81$\pm$0.02 & 1.84$\pm$0.02 & 0.25$\pm$0.02 & 10.7$\pm$0.06 & 2.09$\pm$0.19 & 1.33/133 & 1.86$\pm$0.02 & 25.28$^{+1.92}_{-1.68}$ & 10.09$\pm$0.06 & 1.41/135 \\

T9  & 0.92$\pm$0.03 & 1.85$\pm$0.02 & 0.29$\pm$0.003 & 10.56$\pm$0.06 & 1.89$\pm$0.12 & 1.45/139 & 1.86$\pm$0.02 & 19.56$^{+1.51}_{-1.32}$ & 10.79$\pm$0.06 & 1.84/127 \\

T10 & 0.78$\pm$0.02 & 1.65$\pm$0.02 & 0.41$\pm$0.02 & 12.47$\pm$0.07 & 2.67$\pm$0.18 & 1.89/129 & 1.75$\pm$0.02 & 14.48$^{+0.79}_{-0.72}$ & 12.51$\pm$0.07 & 1.79/133 \\
\hline \\
\end{tabular}} 
\label{table:specfit}
Note: The exponential cutoff power-law seems to show low reduced $\chi^2$ values in comparison to that from Log Parabola model because of higher systematic (4\% instead of 3\%). E$_{s,p}$ values are evaluated from the best-fit Log parabola parameters as described in Appendix  \ref{sec:appndx1}. $\dagger$: The data fits well with powerlaw instead of logparabola model. Also due to poor statistics the cutoff energy is fixed to 13.68 keV as observed during next observations.
\end{table*}
 
\begin{figure}
\includegraphics[width=0.5\textwidth]{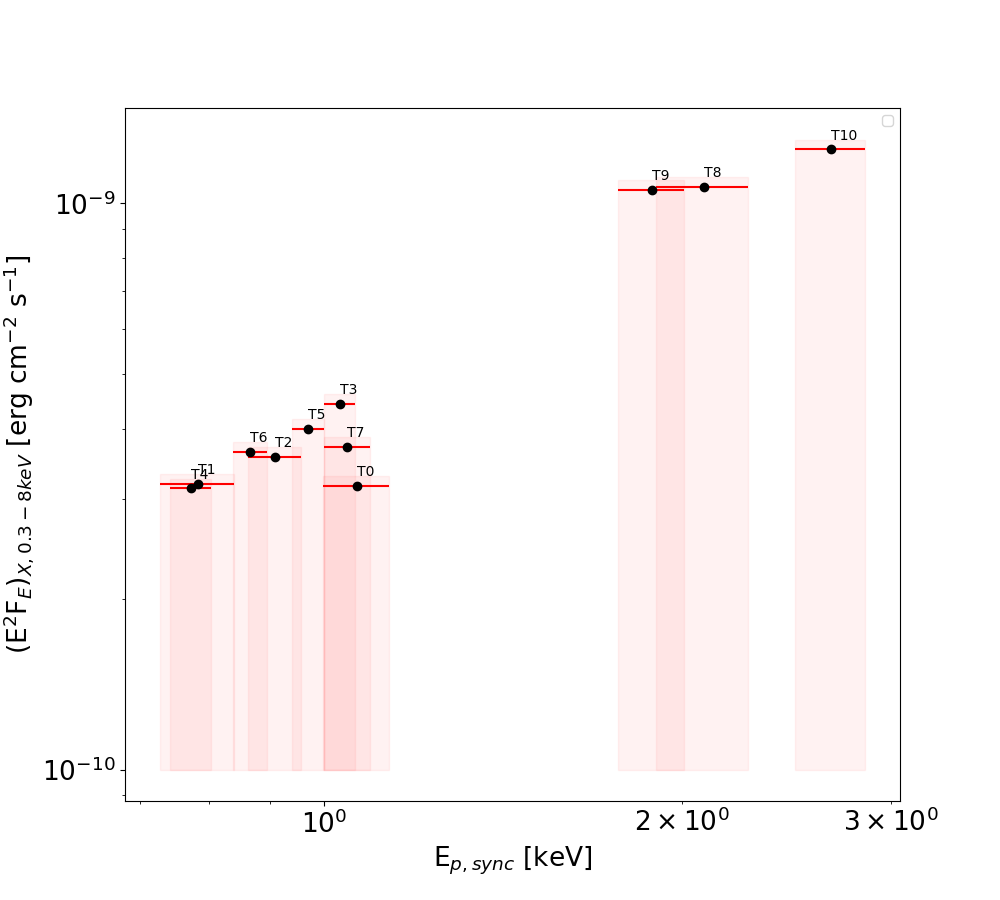}
\caption{\label{fig:peakshiftflux} 
Positions of the synchrotron peaks estimated for the different time bins (T0 \ldots T10) defined in Table \ref{tab:timesegref} and Fig. \ref{fig:zmdlc4long}.}
\end{figure}

Fig. \ref{fig:specfits} clearly establishes prominent flux and spectral variations. The vertical dashed lines show the positions of the synchrotron peaks (E$_{s,p}$) which are calculated using equation \ref{eq:peakpos} and reported in Table \ref{table:specfit}.  The projected synchrotron peak position E$_{s, p, pv}$ is below 0.2 keV. It also represents the faintest state as observed with \astrosat. The spectra observed in 2016 (T0, \ldots,T7) clearly show significant variations in E$_{s, p}$ throughout. Fig. \ref{fig:peakshiftflux} illustrates the shift in the synchrotron peaks for various time bins; see also Table \ref{tab:timesegref}. Within the 1$\sigma$ confidence intervals (Fig. \ref{fig:peakshiftflux}), the shifts in E$_{s, p}$ are correlated to the flux states for the outbursts in 2016 and 2017.  

This investigation establishes that the spectra of \source\ in the 0.3$-$30 keV band become harder with increasing X-ray flux, and subsequently, the peak of the synchrotron emission shifts towards higher energies. The hardening of spectra and shift in E$_{s, p}$ is linked to the particle energization. These motivate us to investigate the relationship between fluxes and E$_{s, p}$ further by utilizing a) modelling SXT spectra sampling smaller portions of the flares and, and b) fitting around 400 XRT spectra observed between January 2015 to February 2021. The spectral modelling of the SXT spectra with such a sampling provides us with a unique independent data set representing the two major outbursts in 2016 and 2017, while the integration of the XRT data, sampling the X-ray variations over 6 years, adds to a general understanding of the above relationship. Additional details are discussed in \S\ref{subsec:corrana}. 

\subsection{Correlation Analysis \label{subsec:corrana}}

\begin{figure*}
\includegraphics[height = 0.55\textwidth, width=0.49\textwidth]{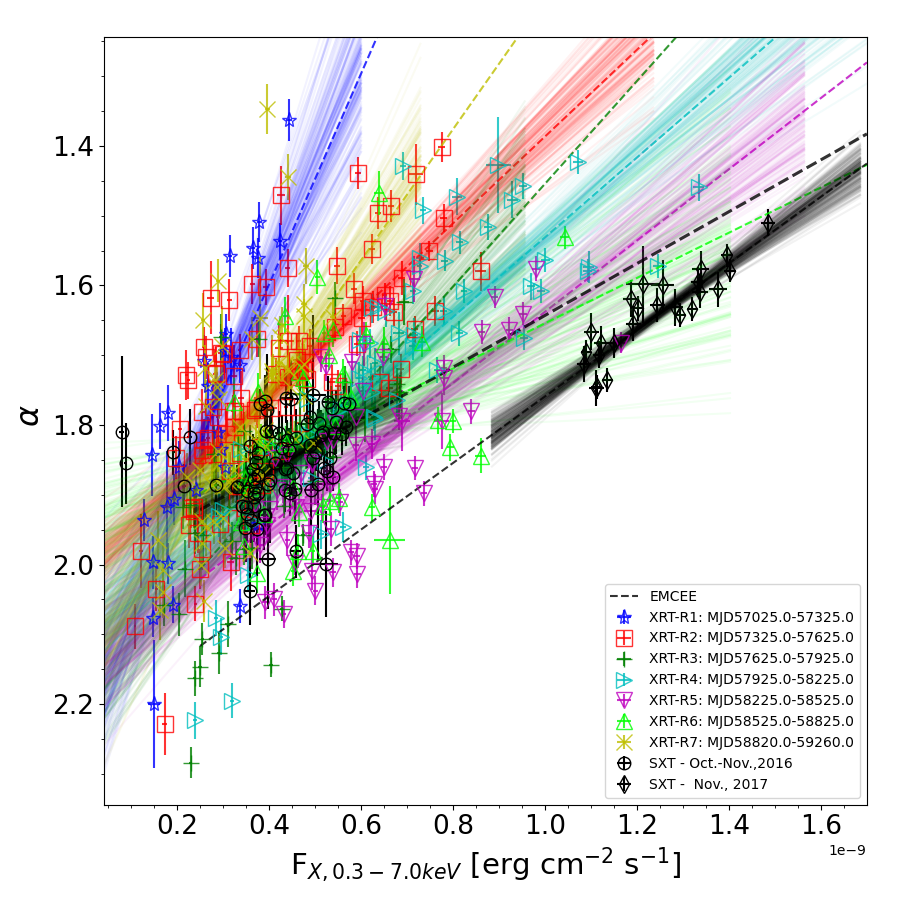}
\includegraphics[ width=0.52\textwidth]{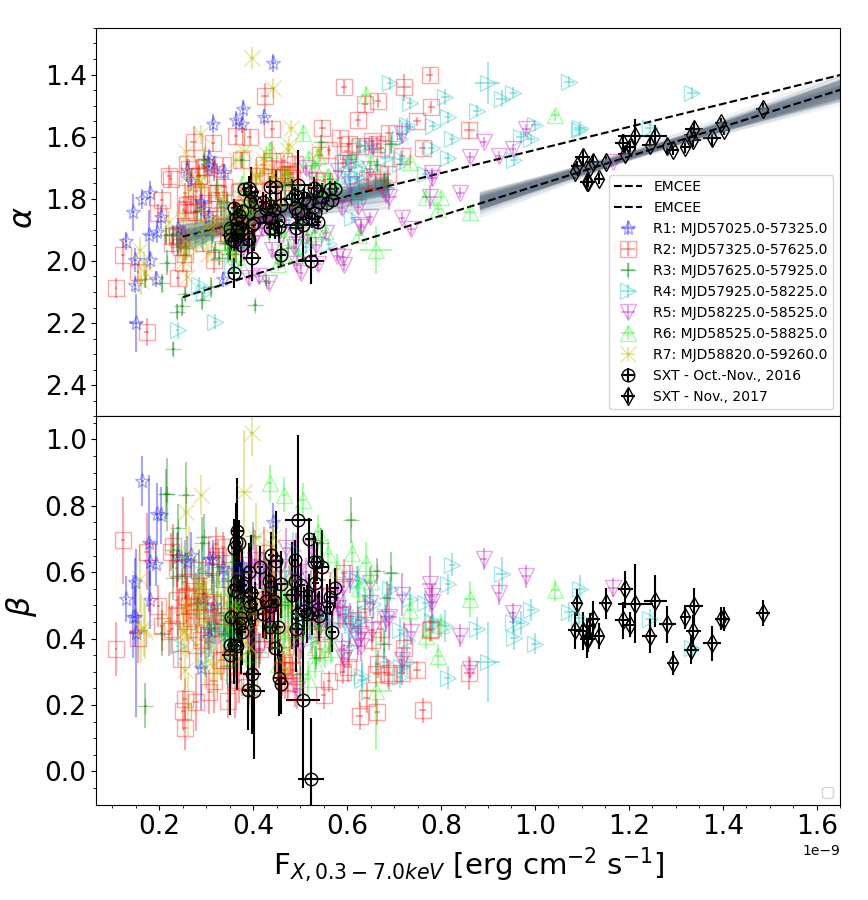}
 \caption{{\footnotesize \label{fig:hardness_intensity}  {\bf Left, hereafter Fig. \ref{fig:hardness_intensity}a:} Spectral slope $\alpha$ vs. integrated X-ray flux in the 0.3-7.0 keV band extracted from the time-resolved spectral modeling of SXT observations and the spectra from individual XRT observations starting from 2015-01-04 to 2021-02-15. The open black circles and open black diamonds refer to the SXT observations in 2016 and 2017, respectively. The other XRT observations are divided into seven time segments(R1,\ldots,R7) and are shown by various symbols and colors as described in the legends. The dashed lines of different colors show the best fit linear regressions using the EMCEE method. The various colored thin solid lines indicate the spread around the best fit lines for a particular group of data.
 {\bf Right, hereafter Fig. \ref{fig:hardness_intensity}b:} Best fit linear correlations with data on $\alpha$ (top panel) and $\beta$ (without any fit --- bottom panel) vs. flux. For clarity, the top panel does not include best fit lines for the various time-groups of XRT observations to illustrate the offset between the  spectral-index dependencies on flux between the flares in 2016 and 2017.
 }}
\label{fig:indxflx}
\end{figure*}

\begin{table*}
\centering
\ra{1.1}
\caption{\label{tab:corrana} Results from correlation studies:}
{\footnotesize \begin{tabular}{@{}llllllllllll@{}}\toprule 
& \multicolumn{2}{c}{$R1 = MJD 57025.0-57325.0$} & \phantom{abc}& \multicolumn{2}{c}{$R2 = MJD 57325.0-57625.0$} & \phantom{abc} & \multicolumn{2}{c}{$R3 = MJD 57625.0-57925.0$}\\
\cmidrule{2-3} \cmidrule{5-6} \cmidrule{8-9} & $r_s$ & $p$   && $r_s$ & $p$  &&  $r_s$ & $p$\\
\cmidrule{2-3} \cmidrule{5-6} \cmidrule{8-9}
$\alpha$ and $F_X$ & $-$0.70 & 1.08$\times 10^{-5}$ && $-$0.72 & 1.45$\times 10^{-17}$ && $-$0.69 & 3.64$\times 10^{-8}$ \\
$E_{s,p}$ and $F_{X}$ &  0.50 & 4.0$\times 10^{-3}$ &&  0.79 & 2.64$\times 10^{-23}$ &&  0.71 & 9.74$\times 10^{-9}$\\
$E_{s,p}$ and  $\alpha$ & $-$0.79 & 1.26$\times 10^{-7}$ && $-$0.85 & 6.02$\times 10^{-29}$ && $-$0.95 & 1.24$\times 10^{-26}$ \\
$\beta$ and $F_{X,0.3-7\rm\,keV}$ & $-$0.11 & 0.55 && $-$0.40 & 3.37$\times 10^{-5}$ && $-$0.25 & 8.19$\times 10^{-2}$ \\
$\alpha$ and  $\beta$ & $-$0.10 & 0.59 && 0.04 & 0.69 && 0.15 & 0.28 \\
$m$ & $-$0.15$\pm$0.03 & -- &&  $-$0.06$\pm$0.006 & -- &&  $-$0.074$\pm$0.01 & --\\
$b$  & 2.20$\pm$0.08& -- &&  2.01$\pm$0.03 & -- && 2.19$\pm$0.05 & --\\
F$_{var}$&  0.37$\pm$0.009 &  && 0.41$\pm$0.001 &  && 0.35$\pm$0.002 & \\ \\

& \multicolumn{2}{c}{$R4 = MJD 57925.0-58225.0$} & \phantom{abc}& \multicolumn{2}{c}{$R5 = MJD 58225.0-58525.0$} & \phantom{abc} & \multicolumn{2}{c}{$R6 = MJD 58525.0-58825.0$}\\
\cmidrule{2-3} \cmidrule{5-6} \cmidrule{8-9} & $r_s$ & $p$   && $r_s$ & $p$  &&  $r_s$ & $p$\\\cmidrule{2-3} \cmidrule{5-6} \cmidrule{8-9}
$\alpha$ and $F_X$ & $-$0.87 & 1.80$\times 10^{-19}$ && $-$0.68 & 1.75$\times 10^{-9}$ && $-$0.37 & 1.43$\times 10^{-2}$ \\
$E_{s,p}$ and $F_{X}$ &  0.84 & 7.55$\times 10^{-17}$ &&  0.74 & 1.65$\times 10^{-11}$ && 0.47 & 1.71$\times 10^{-3}$ \\
$E_{s,p}$ and  $\alpha$ & $-$0.92 & 3.26$\times 10^{-24}$ && $-$0.98 & 6.79$\times 10^{-41}$ && $-$0.89 & 4.34$\times 10^{-15}$ \\
$\beta$ and $F_{X,0.3-7\rm\,keV}$ & $-$0.13 & 0.32 && $-$0.22 & 8.64$\times 10^{-2}$ && 0.02 & 0.92 \\
$\alpha$ and  $\beta$ & 0.05 & 0.71 && $-$0.21 & 0.11 && $-$0.56 & 1.27$\times 10^{-4}$ \\
$m$ & $-$0.057$\pm$0.005 & -- &&  $-$0.05$\pm$0.006 & -- &&  $-$0.03$\pm$0.01& --\\
$b$  & 2.10$\pm$0.04 & -- &&  2.15$\pm$0.04& -- && 1.98$\pm$0.06& --\\
F$_{var}$ & 0.36$\pm$0.002 &  && 0.28$\pm$0.001 &  && 0.30$\pm$0.002 &  \\\\

& \multicolumn{2}{c}{$R7 = MJD 58820.0-59260.0$} & \phantom{abc}& \multicolumn{2}{c}{$AstroSat$ - Oct., Nov., 2016} & \phantom{abc} & \multicolumn{2}{c}{$AstroSat$ - Nov., 2017}\\
\cmidrule{2-3} \cmidrule{5-6} \cmidrule{8-9} & $r_s$ & $p$   && $r_s$ & $p$  &&  $r_s$ & $p$\\\cmidrule{2-3} \cmidrule{5-6} \cmidrule{8-9}
$\alpha$ and $F_X$ & $-$0.63 & 4.13$\times 10^{-8}$ && $-$0.44 & 7.66$\times 10^{-5}$ && $-$0.81 & 4.07$\times 10^{-7}$ \\
$E_{s,p}$ and $F_{X}$ & 0.69 & 3.71$\times 10^{-10}$ && 0.52 & 2.04$\times 10^{-6}$ &&  0.83 & 7.75$\times 10^{-8}$ \\
$E_{s,p}$ and  $\alpha$ & $-$0.90 & 1.09$\times 10^{-23}$ && $-$0.77 & 6.79$\times 10^{-16}$ && $-$0.68 & 9.71$\times 10^{-5}$ \\
$\beta$ and $F_{X,0.3-7\rm\,keV}$ & 0.068 & 0.59 && $-$0.089 & 0.44 && $-$0.11 & 0.58 \\
$\alpha$ and  $\beta$ & $-$0.48 & 7.41$\times 10^{-5}$ && $-$0.47 & 1.88$\times 10^{-5}$ && $-$0.23 & 0.25 \\
$m$ & $-$0.09$\pm$0.01 & -- &&  $-$0.04$\pm$0.009 & -- &&  $-$0.05$\pm$0.004 & --\\
$b$  & 2.13$\pm$0.05& -- &&  2.01$\pm$0.04 & -- && 2.23$\pm$0.05 & --\\
F$_{var}$ & 0.26$\pm$0.002 &  && 0.24$\pm$0.003 &  && 0.087$\pm$0.007 &  \\ 

\bottomrule 
\end{tabular}} \\
{\bf Note:} Here, $r_s$ is the Spearman rank co-efficient and $p$ denotes its null hypothesis probability. 
$m$ and $b$ are respectively, the slope and the intercept of the best-fit EMCEE straight line between $\alpha$ v/s $F_{X,0.3-7\rm\,keV}$. F$_{var}$ is the fractional variability amplitude. 
\end{table*} 
This section describes the detailed correlation analysis between various spectral parameters and other dependent quantities for the time-segments R1,\ldots,R7 and two sets of \astrosat\ observations [Fig. \ref{fig:longlc}], separately.  The $\alpha$ v/s F$_{X,0.3-7keV}$ distribution is fitted with a straight-line and is depicted in the Fig.\ref{fig:hardness_intensity}a. The various pairs of colors and symbols represent the different R segments. The lines are modelled with slope (m) and intercept (b) as free parameters. In order to extract the best fit parameters, the least square regression (LS), maximum likelihood minimization (ML) and  Monte Carlo Markov Chain (MCMC) techniques are used. For the MCMC calculations, the EMCEE tool in python \citep{Foreman_Mackey_2013}, which is an MIT licensed pure-Python implementation of the \citet{GoodmanWeare2010} Affine Invariant MCMC Ensemble sampler, is used. The best fit parameters derived from the LS and ML techniques are used as initial parameters for the MCMC technique. The thin colored lines in Fig. \ref{fig:hardness_intensity} represent sample lines obtained with the MCMC method, whereas thick dashed lines correspond to the best fits derived with the MCMC technique. 

The total of 81 time-resolved spectra extracted from SXT, as described in \S\ref{subsec:timeresspec}, are grouped into two parts. The first part covers the three epochs in 2016 and the second for the observations in 2017. These data sets are also modeled using the same methods as described above. The correlations between best fit spectral parameters ($\alpha$ and $\beta$) v/s X-ray flux (F$_{X,0.3-7keV}$), E$_{s,p}$ v/s  F$_{X,0.3-7keV}$, and $\alpha$ and $\beta$ are explored for all the R segments. The readers are referred to Table \ref{tab:corrana} for the details of the correlation indicators of various pairs from different R segments and respective best fit parameters of the line. 

The index $\alpha$ shows strong correlations with the integrated flux in the 0.3-7.0 keV band for all data segments. It is clear from this figure that most of the time the X-ray spectrum of \source\ gets harder with increasing flux. However, the slopes of the index v/s $F_{0.3-7.0 keV}$ relationships for the different time segments differ quite a lot. Especially, the slopes (m) for R1 and R7 are higher than for the other segments. Interestingly, R1 and R7 sample the increasing and decreasing tails of the Gaussian fit of the long-term X-ray light curve, respectively. 

The ($m$, $b$) pairs, where $m$ and $b$ are slope and intercept for the fit on the $\alpha$--F$_{X, 0.3-7keV}$ plane, for the \astrosat\ observations in 2016 and R3, differ significantly. Whereas, the slope of \astrosat\ observations in 2017 matches the one in R4, though both showing different intercepts. Note that the \astrosat\ observations mainly sample spectra around particular outbursts, while the \swift\ observations are spread over a longer duration and hence represent a general behaviour over long periods.  

The right panel of Fig. \ref{fig:hardness_intensity}b shows the flux dependence of the index (top) and the curvature parameter (bottom). The best fit lines for the R segments are not plotted in the top panel for the sake of clarity. The two sets of \astrosat\ observations clearly represent two spectrally distinct states of variability of \source. As clearly shown in Fig. \ref{fig:hardness_intensity}, the outburst in 2017 does not only exhibit the highest X-ray flux, but also follows an entirely different track in the spectral-index -- flux plane.  

\begin{figure*}
\includegraphics[width=0.49\textwidth]{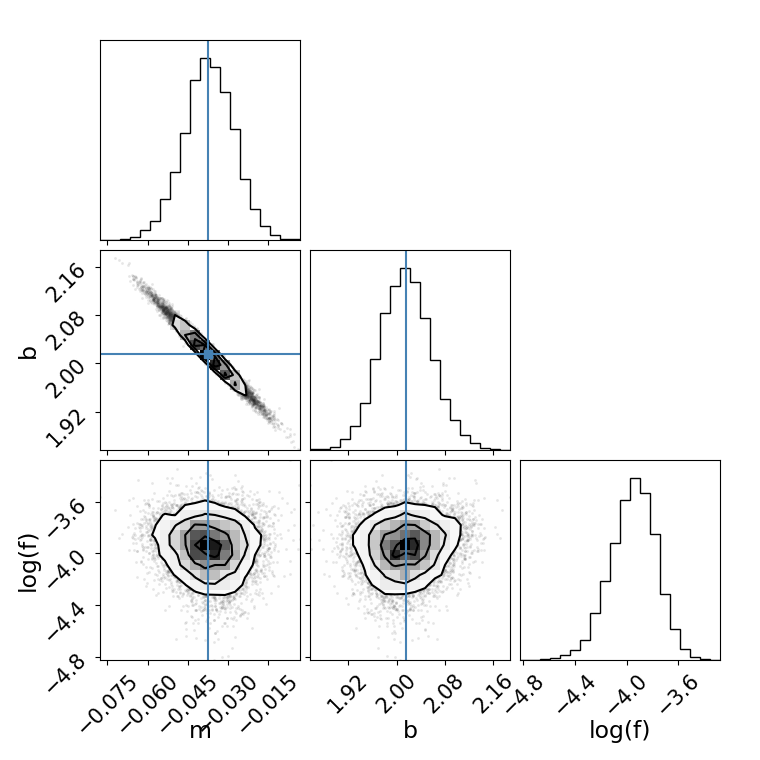}
\includegraphics[width=0.49\textwidth]{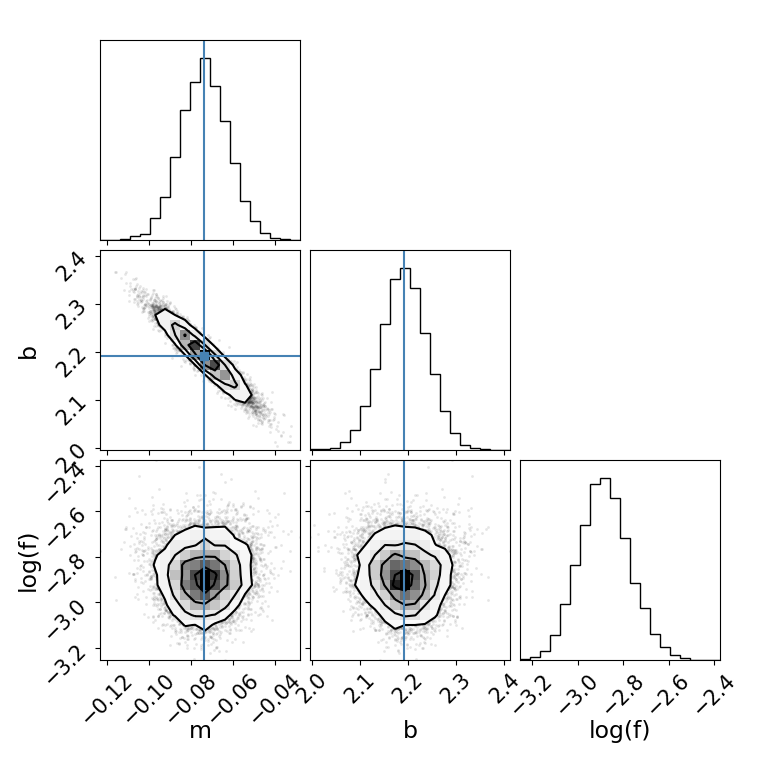}
\caption{{\footnotesize \label{fig:diagplt} Diagnostics plots for the best fit parameters for the $\alpha$ v/s flux correlations. {\bf Left, hereafter Fig. \ref{fig:diagplt}a:} Plot for the time-resolved SXT spectra from the 2016 flare. {\bf Right, hereafter Fig. \ref{fig:diagplt}b:} Plot for all the XRT spectra observed by \swift\ over the period MJD 57026 to MJD 57637.}}
\end{figure*}

Fig. \ref{fig:diagplt} visualizes the confidence intervals and probability distribution for the best fit parameters corresponding to the \astrosat\ observations in 2016 and R3, respectively. The plots are made using the python package {\tt corner}, which corresponds to the MCMC fitting method. These diagnostic plots are indicative of a reasonably good convergence and an acceptable quality of the parameter estimations. 
The correlations between flux enhancements and changes in the spectral parameters $\alpha$ and $\beta$ may provide unique information about the energization of particles responsible for the flux enhancements during a particular flare. Studies by other authors \citep{MAGIC2020, 2018MNRAS.473.2542K, 2018ApJ...858...68K, Krawczynski2004} have shown that for a handful of blazars the curvature and index are inversely correlated with an increase in flux. More specifically, an increase in flux is accompanied by a hardening of the spectrum (decreasing $\alpha$) and a decrease in curvature (decreasing $\beta$). The spectral constraints derived for the \astrosat-SXT and \swift-XRT observations confirm the strong positive correlations between the flux and the hardening of the spectrum. However, owing to the large uncertainties, the curvature parameter $\beta$ (shown in the bottom panel of Fig. \ref{fig:hardness_intensity}b) does not show a significant correlation with the flux. 

These spectral changes result in a shift of the peak position of the X-ray part of the SED and hence in the position of the synchrotron peak. Such a shift of the synchrotron peak towards higher energies with increasing integrated flux is seen in most HBLs and is equivalent to the ``bluer-when-brighter" trend seen in the optical continuum spectra of many BL Lac objects \citep{2010ApJ...725.2344B, 2011AJ....141...65D, 2018A&A...619A..93B, Kaur2017}. The best fit $\alpha$ and $\beta$ values are used in equation \ref{eq:peakpos} to derive the position of the peak of the synchrotron component. 

\begin{figure}

\includegraphics[width=0.5\textwidth]{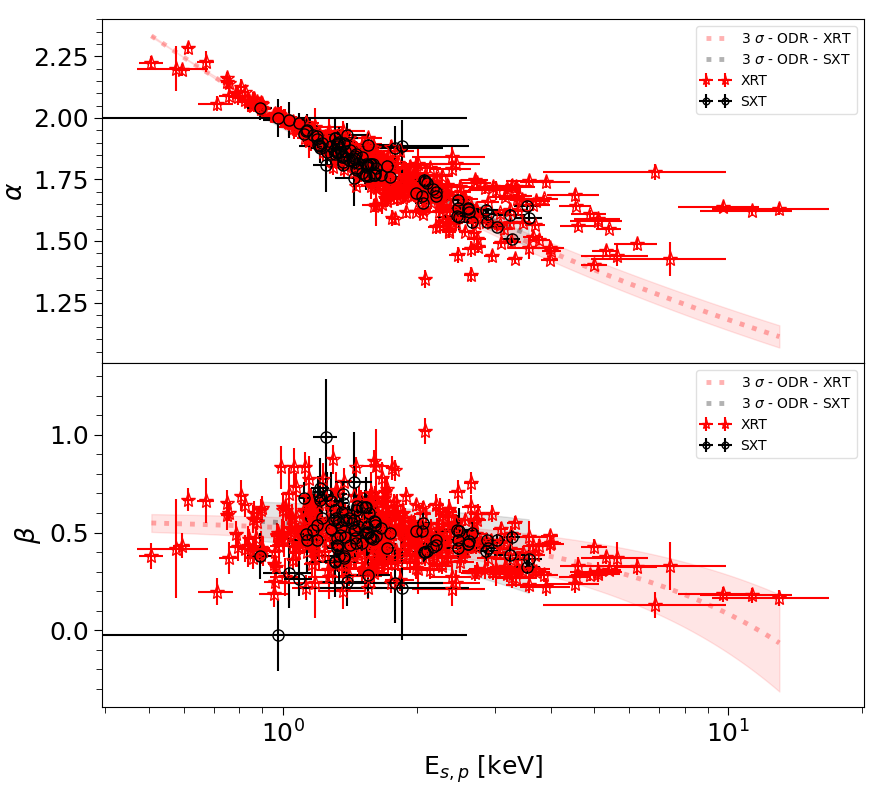}
\includegraphics[width=0.5\textwidth]{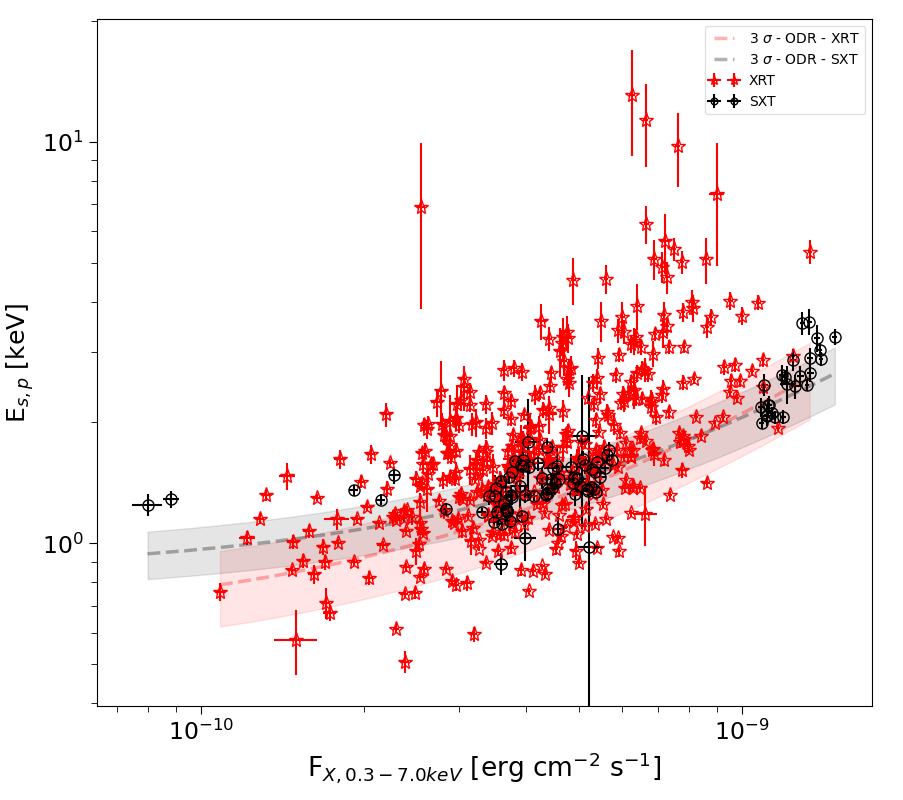}

\caption{{\footnotesize \label{fig:peakFlux} Synchrotron Peak Position and Unabsorbed X-ray flux.{ \bf Top, hereafter Fig. \ref{fig:peakFlux}a:} Dependence of the spectral slope ($\alpha$; top panel) and the curvature ($\beta$; bottom panel) on the position of the synchrotron peak (E$_{s,p}$). The filled red stars represent the XRT data whereas the measurements from the SXT are shown by filled black circles. {\bf Bottom, hereafter Fig. \ref{fig:peakFlux}b:} Correlations between the estimated synchrotron peak positions and the integrated X-ray flux measured in the 0.3-7.0 keV band.The dotted red and black lines represent the best fit linear correlations for XRT and SXT observations, respectively. The shaded areas around the best fit curves in both the figures represents 3$\sigma$ confidence interval.}}
\end{figure}

Fig. \ref{fig:peakFlux}a shows the relationship between the synchtrotron peak energy (E$_{s,p}$) and the spectral parameters $\alpha$ and $\beta$, respectively. The index parameter $\alpha$ is well correlated with the synchrotron peak position above an energy of $\sim 0.5$~keV, while such a statement cannot be made for the curvature parameter $\beta$.  

Fig. \ref{fig:peakFlux}b shows the relationship between E$_{s,p}$ and the integrated flux F$_{X,0.3-7.0 keV}$, confirming the trend of the synchrotron peak shifting to higher energies with increasing flux. While the relationship between E$_{s,p}$ v/s $\alpha$ is well represented by a power-law function, the relationship between F$_{X, 0.3-7.0 keV}$ and E$_{s,p}$ is reasonably fitted with a linear function (All XRT: slope=1.47$\pm$0.11, intercept=0.63$\pm$0.04; All SXT:slope=1.21$\pm$0.07, intercept=0.84$\pm$0.05). In order to further test the correlations, we have performed a Spearman rank correlation analysis of these observable parameters for the spectra obtained during R1,\ldots,R6, and the results are reported in Table \ref{tab:corrana}.

\subsection{SEDs, Modeling and Interpretation \label{sec:modeling}}

The X-ray light curves shown in Fig. \ref{fig:mwllc} indicate the presence
of clearly discernible, individual flares, both in 2016 and 2017. We interpret 
these events as the result of mildly relativistic shocks propagating through 
the jet of \source. In order to model the light curves and SEDs during the 
period of our \astrosat\ observations in 2016 and 2017, we employ the time-dependent 
shock-in-jet model of \cite{BB19}. In this model, hybrid thermal + non-thermal 
electron distributions are generated via a Monte-Carlo simulations \citep{SB12} 
of diffusive shock acceleration by a mildly relativistic, oblique shock. As 
a representative choice of shock parameters, we assume a shock speed of $v_s = 
0.71$~c (in the co-moving frame of the jet material), a magnetic-field
obliquity of $\Theta_{\rm Bf1} = 32^o.3$, an up-stream gas temperature of 
$5.45 \times 10^7$~K, and a compression ratio of $r = 3.71$ \citep[see][for 
a motivation and discussion of these choices]{BB19}. The Monte-Carlo simulations
of diffusive shock acceleration parameterize the electrons' mean free path to
pitch-angle scattering as $\lambda_{\rm pas} = \eta_0 \, r_g \, p^{\alpha - 1}$
where $r_g$ is an electron's Larmor radius and $p$ its momentum. $\eta_0$ and $\alpha$
are free parameters in the simulation. Since $r_g \propto p$, the mean free path 
scales as $\lambda_{\rm pas} \propto p^{\alpha}$. 

The time-dependent radiative output from the resulting hybrid electron distributions 
is evaluated using the radiation transfer schemes of \cite{BC02,Boettcher13}. In addition 
to $\eta_0$ and $\alpha$, free parameters of the model are the shock-dissipated power 
transferred to relativistic electrons (termed ``injection luminosity'', $L_{\rm inj}$, 
in the following), the magnetic-field strength $B$, the bulk Lorentz factor $\Gamma$
of the emission region's propagation along the jet, the viewing angle $\theta_{\rm obs}$,
and the radius $R$ of the emission region, which is assumed spherical for the purpose of 
the radiation-transfer simulations. $L_{\rm inj}$, $B$, and $R$ are defined in the 
co-moving frame of the emission region, while $\theta_{\rm obs}$ is the viewing angle 
in the observer's frame. 

Since we find a satisfactory fit to snap-shot SEDs and light curves of 1ES~1959+650 with 
synchrotron and synchrotron-self-Compton (SSC) as the dominant radiation mechanisms (i.e., 
a pure leptonic SSC model), we do not consider any contribution from putative external 
radiation fields to the target photon field for Compton scattering to produce the 
$\gamma$-ray emission. 

\begin{figure*}
\includegraphics[width=0.49\textwidth]{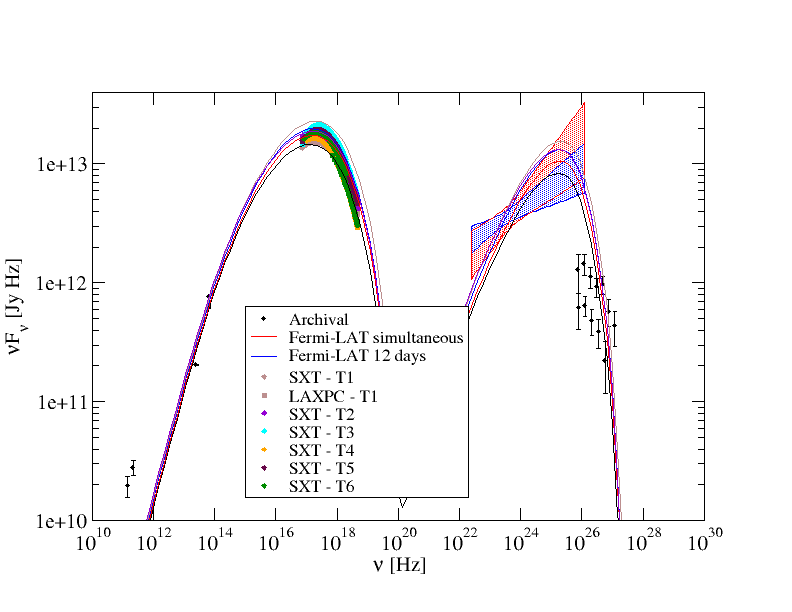}
\includegraphics[width=0.49\textwidth]{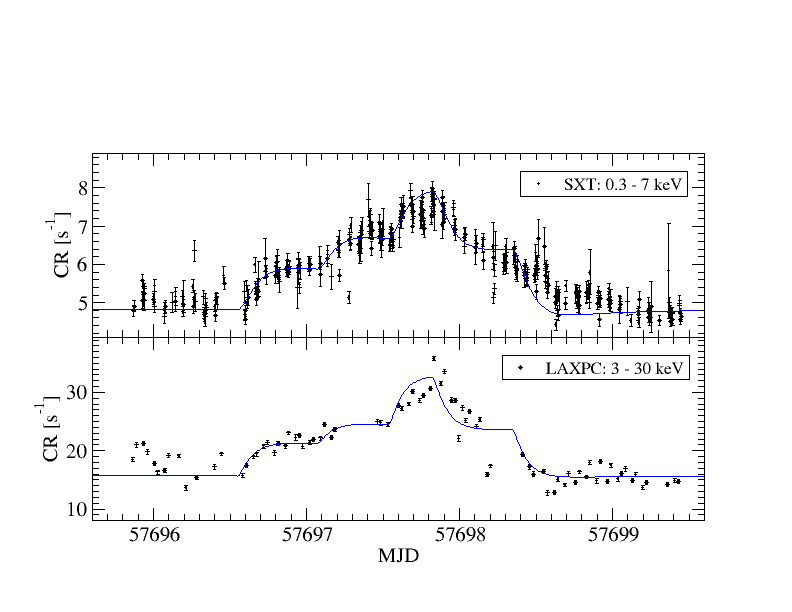}\\
\includegraphics[width=0.49\textwidth]{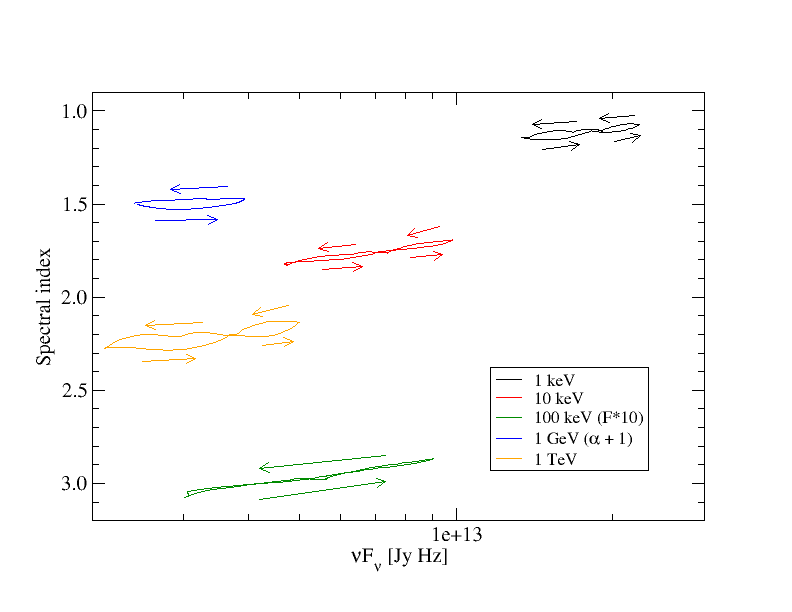}
\includegraphics[width=0.49\textwidth]{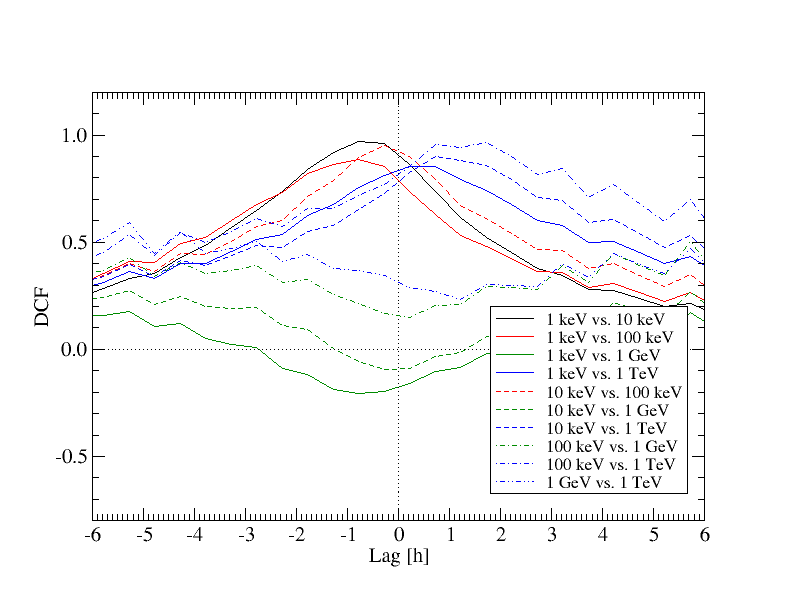}
\caption{\label{model2016} Model fits to SEDs and light curves of 1ES~1959+650 during the \astrosat\ observations in 2016. {\bf Top left (a):} Snap-shot SEDs at various times during the simulation; {\bf top right (b):} Model fits to the \astrosat\ light curves; {\bf bottom left (c):} Predicted hardness-intensity correlations in various energy 
bands. Note that for plotting on the same scale, the 100 keV curve has been shifted up in flux (i.e., right) by a factor of 10, while the 1 GeV curve has been shifted up in spectral index (i.e., down in the plot) by 1; {\bf bottom right (d):} Predicted discrete correlation functions between light curves in various energy bands. }
\end{figure*}

\begin{figure*}
\includegraphics[width=0.49\textwidth]{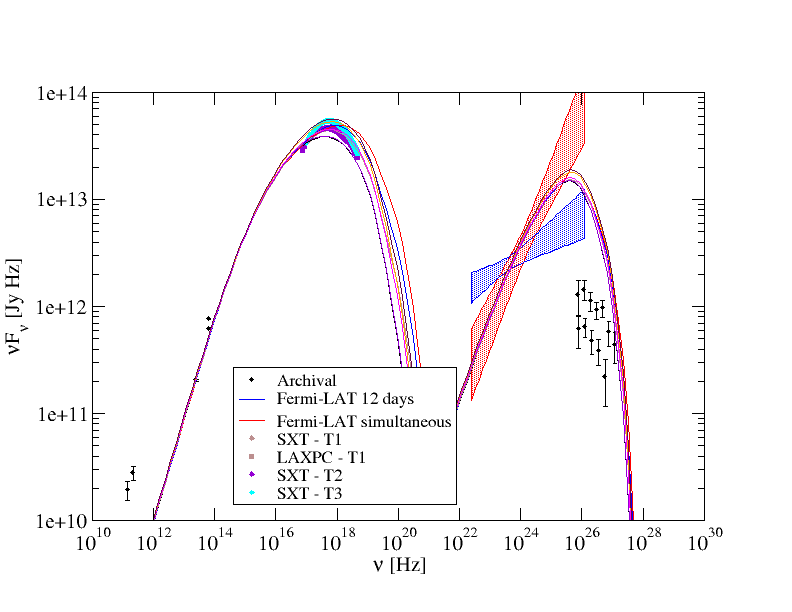}
\includegraphics[width=0.49\textwidth]{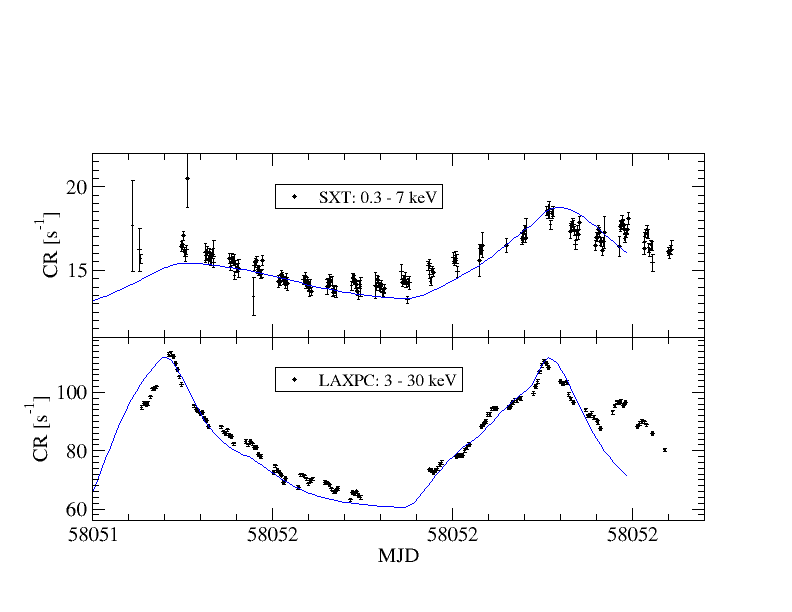}\\
\includegraphics[width=0.49\textwidth]{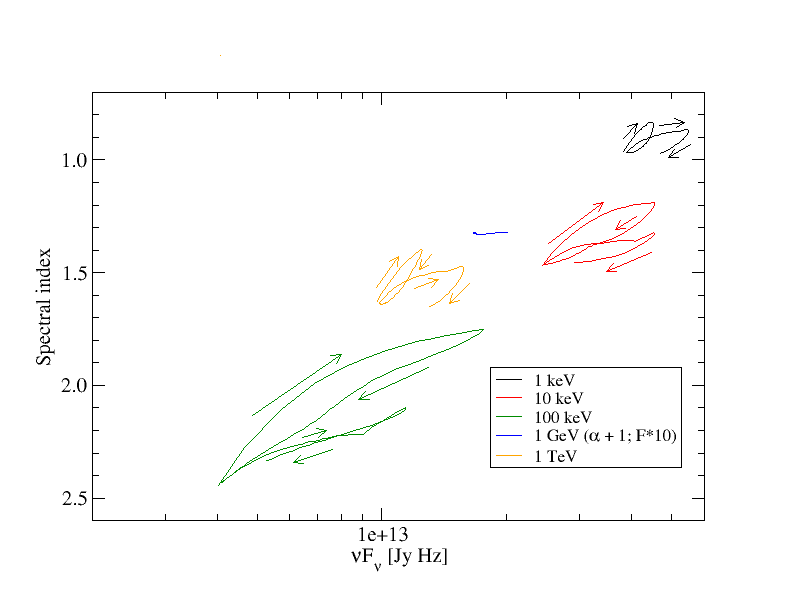}
\includegraphics[width=0.49\textwidth]{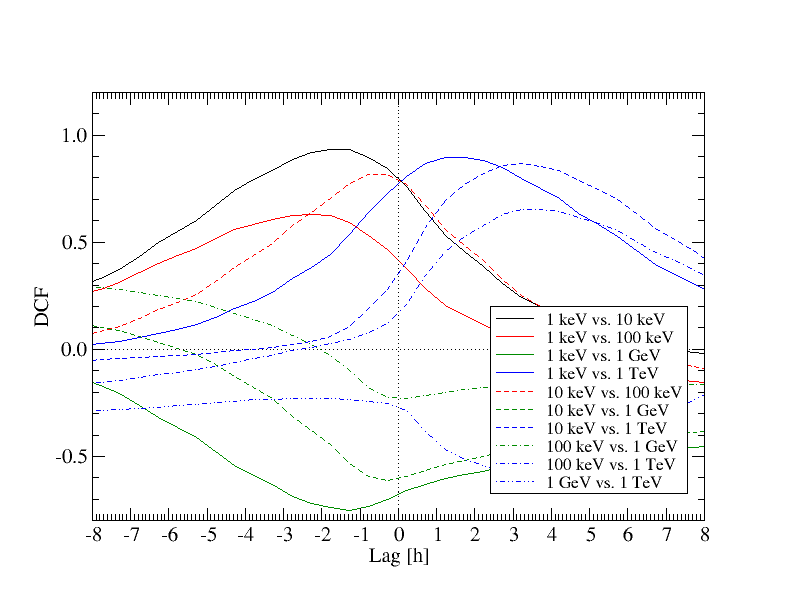}
\caption{\label{model2017} Same as Fig. \ref{model2016}, but for the \astrosat\ observations in 2017. Note that in the panel c (bottom left), the 1 GeV hardness-intensity curve has been shifted up (i.e., right) in flux by a factor of 10 and up in spectral index by 1 (i.e., down in the plot) for plotting on the same scale as the other curves. }
\end{figure*}

Our fitting procedure starts out with a quiescent-state configuration, reproducing the 
low-state SED of 1E~1959+650, shown by the black model curves in Figs. \ref{model2016}a 
and \ref{model2017}a, with model parameters listed in Tab. \ref{quiescent_parameters}.

\begin{table}
\caption{\label{quiescent_parameters}Parameters of our model fits to the quiescent-state SEDs of 1ES~1959+650 in 2016 and 2017, respectively.}
\begin{tabular}{ccc}
\hline
Parameter [units] & 2016 & 2017 \cr
\hline
$\eta_0$ & 60 & 40 \cr
$\alpha$ & 1.9 & 1.8 \cr
$L_{\rm inj}$ [erg/s] & $2.5 \times 10^{40}$ & $2.8 \times 10^{40}$ \cr
$B$ [G] & 0.15 & 0.08 \cr
$\Gamma$ & 20 & 20 \cr
$\theta_{\rm obs}$ [deg] & 2.87 & 2.34 \cr
$R$ [cm] & 6.e15 & 1.e16 \cr
\hline
\end{tabular} 
\end{table}

The zoomed-in light curves in Figs. \ref{model2016}b and \ref{model2017}b clearly suggest that the observed X-ray flaring behaviours both in 2016 and 2017 cannot be modelled by one single, impulsive particle acceleration event, but they are indicative of a 
succession of several shocks throughout the emission region. Specifically, in order to find a satisfactory representation of both the SXT and the LAXPC light curves during 2016, a succession of 4 shocks of different strength is required. All shocks are characterized by an increased injection luminosity and a global decrease of the 
pitch-angle mean free path, parameterized by a slightly smaller value of $\eta_0$. A consequence of this change is more efficient particle acceleration to higher energies, resulting in the observed larger variability amplitude in the LAXPC (3 -- 30~keV) band compared to the SXT (0.3 -- 7~keV) band. The parameters adopted for the 4 shocks in our 2016 simulation are listed in Table \ref{2016shocks}. 
Representative snap-shot SEDs are shown in Fig. \ref{model2016}a, and the LAXPC and SXT light curves and model fits are presented in Fig. \ref{model2016}b. Since both the UV and {\it Fermi}-LAT $\gamma$-ray light curves consist of only very few points (and upper limits), they are not constraining for our fits and are not shown in the figure. We have, however, verified that our model predictions are consistent with the data. Figs. \ref{model2016}c and \ref{model2016}d show the 
predicted spectral hysteresis in a hardness-intensity diagram at various photon energies, and the predicted cross-correlations between the light curves at different photon energies, respectively. 

\begin{table}
\caption{\label{2016shocks}Parameter variations for the fits to MWL light curves and SEDs in 2016.}
\begin{tabular}{cccc}
\hline
Parameter [units] & $L_{\rm inj}$ [erg/s] & $\eta_0$ & $\alpha$ \cr
\hline
Quiescence & $2.5 \times 10^{40}$ & 60 & 1.9 \cr
Shock 1 & $3.0 \times 10^{40} $   & 50 & 1.9 \cr
Shock 2 & $3.5 \times 10^{40}$    & 50 & 1.9 \cr
Shock 3 & $4.1 \times 10^{40}$    & 40 & 1.9 \cr
Shock 4 & $3.4 \times 10^{40}$    & 50 & 1.9 \cr
\hline
\end{tabular}
\end{table}

The observed hardness-intensity correlations plotted in Fig. \ref{fig:hardness_intensity} (left) indicate a systematic offset by a factor of $\sim 2$ in flux of the data from 2017 with respect to the 2016 data in flux. However, within the individual yearly data sets, they show consistent harder-when-brighter trends. Such a systematic offset may be explained by a slight change of the Doppler factor (by a factor of $\sim 1.2$) without any changes in the underlying particle-acceleration and emission physics. A slight change of the viewing angle (from $\theta_{\rm obs} = 2.^o87$ in 2016 to $2.^o34$ in 2017) is sufficient to reproduce such a change in Doppler factor (from 20 to 24). 
This forms the basis of our modeling of SEDs and X-ray light curves in 2017. Only slight modification of other model parameters are required for our quiescent-state fit to the 2017 SEDs (see Tab. \ref{quiescent_parameters}).

The X-ray light curves (Fig. \ref{model2017}b) from the 2017 \astrosat\ observations show two clearly distinct major flaring episodes near the beginning and the end of the observations. Clearly, again, multiple shocks are required in order to provide a satisfactory representation of these complex light curves. Specifically, we find 
a good match with the SXT and LAXPC adopting a succession of 5 shocks, all characterized by a larger injection luminosity compared to the quiescent state as well as more efficient particle acceleration to higher energies. The latter is primarily achieved 
through a change in the mean-free-path parameter $\eta_0$. The parameters adopted for the 5 subsequent shocks are listed in Tab. \ref{2017shocks}. 

\begin{table}
\caption{\label{2017shocks}Parameter variations for the fits to MWL light curves and SEDs in 2017.} 
\begin{tabular}{cccc}
\hline
Parameter [units] & $L_{\rm inj}$ [erg/s] & $\eta_0$ & $\alpha$ \cr
\hline
Quiescence & $2.8 \times 10^{40}$ & 40 & 1.8 \cr
Shock 1 & $3.5 \times 10^{40} $   & 30 & 1.7 \cr
Shock 2 & $3.0 \times 10^{40}$    & 30 & 1.8 \cr
Shock 3 & $3.6 \times 10^{40}$    & 25 & 1.8 \cr
Shock 4 & $4.3 \times 10^{40}$    & 25 & 1.8 \cr
Shock 5 & $5.1 \times 10^{40}$    & 15 & 1.8 \cr
\hline
\end{tabular}
\end{table}

Given the relative simplicity of our model setup and the complexity of the SXT and LAXPC light curves both in 2016 and 2017, the correspondence between the observed X-ray light curves and our model predictions is remarkable. For both years, we reproduce the observed general harder-when-brighter trend (see Fig. \ref{fig:hardness_intensity}) with only very moderate spectral hysteresis, which is smaller than the error bars on the SXT and LAXPC flux and spectral-index points, especially for 2016. 
The model 1 keV vs. 10 keV time lags may be considered as a proxy to the observed trends between SXT and LAXPC. For 2016, our model predicts a lag of $\sim 1$~hr of the SXT (1 keV) light curve behind the LAXPC (10 keV) one, while for 2017, this lag is predicted to be slightly larger, at $\sim 1.5$~hr. Predicted lags between X-ray 
and $\gamma$-ray bands are equally of the order of $\lesssim 1$~hr. \label{subsec:results}
\section{Discussion and Summary \label{discussSum}}

Our analysis, presented above, clearly establishes that \source\ underwent significant flaring activity in X-rays over the 6 years during 2015 to 2021 (see Fig. \ref{fig:longlc}). The overall long-term X-ray variability over these years shows a profile that is symmetric and well represented by a broad (FWHM$\sim$785 days) Gaussian function peaking around April 2018.  Such a long-term symmetric flux variation is suggestive of a change in the viewing angle, and hence in the Doppler boosting, as the main driver. Most probably the inner jet is curved. Unfortunately, the data coverage is not sufficient to test the jet-precession theory, which predicts periodic changes in the flux. The X-ray light curves show that there are many short-time-scale variations superimposed on the long-term symmetric variations throughout (Fig. \ref{fig:zmdlc4long}). 

The detailed light curves observed with \astrosat\ cover different prominent X-ray flares and exhibit shorter-time-scale variations (see Table \ref{tab:tvar} for time-scales and the variability amplitudes.)  During the X-ray outburst in 2016 the source also exhibits noticeable activity at $\gamma$-ray energies, even though no clear correlation pattern emerges. Most importantly, V2-SF1 is accompanied by simultaneous GeV/TeV activity, whereas V2-SF2 (total span of $\sim$ 2 days) seems to be a pure orphan X-ray flare. 

In 2017 \source\ underwent several prominent X-ray outbursts including the one observed with \astrosat, which seems to have a twin-flare-like profile within $\sim$ 2 days (see Fig. \ref{fig:zmdlc4long}b). During this period, \source\ broke its historical X-ray flux record (in the 0.3$-$7.0 keV band) reaching a new maximum, which has not been exceeded until the time of writing (see Fig. \ref{fig:longlc}). The observed X-ray flux showed two high-amplitude, short-time-scale flares without any counterpart at other wavelengths. Therefore, the variations observed with \astrosat\ are again orphan. Interestingly, a few days prior to the X-ray observations prominent variability took place at $\gamma$-ray energies. Unfortunately, the lack of multiwavelength observations during that time precludes any correlation analyses. 

The \astrosat\ observations of the X-ray activity periods in 2016 and 2017 are unique as these provide such a detailed variability profiles for \source\ for the first time. The light curves in the 0.3-7.0 keV band are highly correlated with the ones in the 3.0-30 keV band for all the epochs in 2016 and 2017. 

However, during all epochs, the variability amplitude is larger in the hard X-ray band compared to the soft X-rays. 
The shortest variability time-scale ($\Delta$t; here characterized by the doubling or halving time scales) may be used to calculate an upper limit on the size of the emission region using the light-travel-time argument, R $\leq$  $\frac{c \delta {\Delta}t}{(1+z)}$. The smallest inferred limit on the size of the emission region for the activity periods in 2016 and 2017 are 2.9 mpc and 7.4 mpc, respectively. 

The X-ray spectral investigations reveal significant changes in the spectral shapes for different flares and also for the different segments of the \astrosat\ observations (T0, T1,\ldots,T10). The spectral changes follow a harder-when-brighter trend as is typical for many blazars. The broad-band X-ray spectra (0.3-30 keV) are best represented by a log-parabola model. A similar investigation using time-resolved spectra from SXT and long-term observations from XRT provide similar trends. However, different sets of flares (R1,\ldots,R7) show a slightly different relation between flux and spectral index: Different R's populate significantly different tracks in the $\alpha$ -F$_{X,0.3-7keV}$ plane. The R-segments representing the the beginning (R1) and end (R7) of our data set, on the other hand, show extreme slopes (See Table \ref{tab:tvar} for details). 

The time-resolved spectroscopy of the \astrosat\ observations reveals a strong correlation between the slope ($\alpha$) and the flux (F$_{X,0.3-7.0 keV}$) for both observation periods in 2016 and 2017, similar to the enveloping R-segments (R3 and R4, respectively). However, the tracks in the $\alpha$ v/s F$_{X,0.3-7.0 keV}$ plane of the \astrosat\ data sets are significantly different from the corresponding R-segments. The best fit lines of $\alpha$ v/s F$_{X,0.3-7.0 keV}$ for the two sets comprise significantly different tracks indicating that the spectrum exhibits stronger hardening with increasing flux in 2017 compared to 2016 (see Table \ref{tab:tvar}). This indicates that the "bluer-when-brighter" trend was stronger in 2017 than in 2016, which is also supported by the strong, positive correlation of the synchrotron peak energy ($E_{s,p}$) with the X-ray flux. 

The various X-ray SEDs, namely T1 to T6 from the flare in 2016 and T8 to T10 from the flare in 2017, are combined to generate two broad-band SEDs. A model based on diffusive shock acceleration by mildly relativistic shocks in the jet of \source\ is able to simultaneously reproduce those snap-shot SEDs and the \astrosat\ light curves of both flaring episodes in 2016 and 2017. In this model, multi-wavelength flares are caused by shock-generated turbulence, leading to a reduction of the electrons' mean free path to pitch-angle scattering ($\lambda_{\rm pas}$) and, thus, more efficient particle acceleration.  The instantaneous interplay between shock acceleration and self-consistent radiative (synchrotron, SSC, and external Compton) cooling of the particles in the emission region results in characteristic flux and spectral variability patterns, consistent with the observed ones.  The different flux states between the two flares are well reproduced by a change in the Doppler factor, mediated by a slight change of the viewing angle ($\Delta\theta$ $\sim$ 0.5$^\circ$) within $\sim$ 1 year, and a reduction of the magnetic field.

 This manuscript highlights the flux evolution of various X-ray flares observed between December 2015 and February 2021, their spectral properties, and their correlations. Additionally, a time-dependent leptonic model indicates the particle acceleration responsible for the two X-ray flares in 2016 and 2017 covered in detail with \astrosat. This work additionally avails a huge dataset to the community with the spectral parameters spanning 6 years of X-ray monitoring of {\source}. While a detailed study of the spectral parameters of the $\gamma$-ray bands, as well as their correlation with the X-ray bands, is not possible with the currently available data, such a study may shed light on the emission mechanisms during flaring activities.
The future MeV/GeV/TeV facilities like AMIGO and CTA, shall provide key data for the such investigations.

\acknowledgments
The work of M. B\"ottcher and S. Chandra is supported by the South African
    Research Chairs Initiative (grant no. 64789) of the Department of
    Science and Innovation and the National Research Foundation
    \footnote{Any opinion, finding and conclusion or recommendation in this
    material is that of the authors and the NRF does not accept any
    liability in this regard.} of South Africa. PG acknowledges the
    financial support of Indian Space Research Organisation (ISRO) under
    \astrosat\ archival Data utilization program. The Fermi Science Support
    Center (FSSC) team is acknowledged for providing the data and analysis
    tools. The authors acknowledge the NASA's primary data archive `HEASARC'
    to avail the data form \swift. This research has used the data of
    AstroSat mission of the Indian Space Research Organi-sation (ISRO),
    archived at the Indian Space Science DataCentre (ISSDC). The authors
    acknowledge the proposers and PIs of the dedicated followup programs and
    ToO observations of \source\ using \swift\, without that such a detailed
    dataset would have not possible. The authors would like to acknowledge
    the support from the LAXPC Payload Operation Center(POC) and SXT POC at
    the TIFR, Mumbai for providing supporting data reduction. LaxpcSoft
    software is used for analysis. This work has been performed utilizing
    the calibration data-bases and auxillary analysis tools developed,
    maintained and distributed by AstroSat-SXT team with members from
    various institutions in India and abroad.
\facilities{AstroSat(SXT and LAXPC), Swift(XRT and UVOT), FACT, MIRO}

\software{%
\texttt{astropy} \citep{2013A&A...558A..33A}, 
\texttt{emcee}\footnote{https://emcee.readthedocs.io/en/stable/} \citep{2013PASP..125..306F,2019JOSS....4.1864F}, 
\texttt{extinction}\footnote{https://github.com/kbarbary/extinction} \citep{barbary_kyle_2016_804967},%
\texttt{extinctions}\footnote{https://pypi.org/project/extinctions/}, 
\texttt{fermipy}\footnote{http://fermipy-readthedocs.io/en/latest/} \citep{2017ICRC...35..824W},%
\texttt{heasoft}-v 6.25\footnote{https://heasarc.gsfc.nasa.gov/lheasoft/download.html} \citep{2014ascl.soft08004N},%
\texttt{laxpcsoft}\footnote{(http://astrosat-ssc.iucaa.in/?q=laxpcData)}, 
\texttt{lmfit}\footnote{https://lmfit.github.io/lmfit-py/} \citep{2014zndo.....11813N},%
\texttt{SExtractor} \citep{1996A&AS..117..393B},
\texttt{sxtARFModule}\footnote{https://www.tifr.res.in/~astrosat\_sxt/dataanalysis.html}
          }

\newpage

\appendix

\section{Appendix - I; Synchrotron Peak Frequency}\label{sec:appndx1}
The functional form of the mathematical model used to fit the X-ray spectra of \source\ is given by :

\begin{equation}
    \frac{dN}{dE} = N_0 \left (\frac{E}{E_0} \right )^{-\left [ \alpha + \beta \times \log_{10}\left ( \frac{E}{E_0} \right ) \right ]}
    \label{eq:logpar}
\end{equation}
where E$_0$ is known as pivot energy, $\alpha$ is the photon index and $\beta$ is the curvature parameter of the model. 
The equivalent equation valid for fitting the $\nu$F$_\nu$ plot i.e. X-ray part of the SEDs

\begin{equation}
    \nu F_\nu = N_0 \left (\frac{\nu}{\nu_0} \right )^{\left [ 2 - \alpha - \beta \times \log_{10}\left ( \frac{\nu}{\nu_0} \right ) \right ]}
    \label{Eq:nufnu}
\end{equation}

It is well established that in most of the flux states the synchrotron peak of the broad-band SED of \source lies between 0.3 to 10.0 keV. Hence it is most likely that the maxima of equation \ref{Eq:nufnu} refers to the synchrotron peak energy. That is, $\nu_{syn, peak}$ or $\nu_{s,p}$ [E$_{s,p}$] corresponds to the solution of equation $\frac{d(\nu F_\nu)}{d\nu}$ = 0 
\begin{equation}
    \frac{d(\nu F_\nu)}{d\nu} = N_0 (2-\alpha)\left(\frac{\nu}{\nu_0}\right)^{(1-\alpha)} \left(\frac{\nu}{\nu_0}\right)^{-\beta \times \log_{10}(\frac{\nu}{\nu_0})}  - N_0 \left(\frac{\nu}{\nu_0}\right)^{(2-\alpha)} \times \left (\frac{ 2\beta\times \log_{10} \left(\frac{\nu}{\nu_0}\right) }{\nu} \right) \times \left(\frac{\nu}{\nu_0}\right)^{-\beta \times \log_{10}(\frac{\nu}{\nu_0})} 
\end{equation}

i.e.,

\begin{equation}
    \frac{d(\nu F_\nu)}{d\nu} = N_0 \left[(2-\alpha) - 2\beta \times \log_{10}\left(\frac{\nu}{\nu_0}\right) \right]\times \left(\frac{\nu}{\nu_0}\right)^{[1-\alpha-\beta\times\log_{10}\left(\frac{\nu}{\nu_0}\right)]}
\end{equation}

This means
$\frac{d(\nu F_\nu)}{d\nu}$ = 0 $=>$

\begin{equation}
    \nu_{s,p} =  \nu_0 \times 10^{ (\frac{2-\alpha}{2\beta} )} \,\,\, or ;\,\,\,  E_{s,p} = E_0 \times 10^{ (\frac{2-\alpha}{2\beta} )}
    \label{eq:peakpos}
\end{equation}
here, $\nu_0$ corresponds to the frequency of the photons at 1 keV i.e., $\nu_0$ = 2.4180 $\times$ 10$^{17}$ Hz. Therefore, for a given index $\alpha$ and curvature $\beta$ we can estimate the position of the synchrotron peak using equation \ref{eq:peakpos}. The same formalism is derived as equation 3 in \citet{2004A&A...422..103M}.

\bibliography{new.ms}{}

\begin{center}
\begin{longtable}{|lllllccc|}
\caption[\astrosat  data and the best fit parameter corresponding the model : M1. The complete table is available in a machine-readable format.]{\astrosat\  data and the best fit parameter corresponding the model : M1. The complete table is available in a machine-readable format in the published journal or via email on request.} \label{tab:astspec_par} \\

\hline \multicolumn{1}{|c|}{\textbf{ObsID}} & \multicolumn{1}{|c|}{\textbf{Time}} & \multicolumn{1}{c|}{\textbf{Exp.}} & \multicolumn{1}{c|}{\textbf{\large $\alpha$}} & \multicolumn{1}{c|}{\textbf{\large $\beta$}} & \multicolumn{1}{c|}{\textbf{$\chi^2_\nu$/dof}} & \multicolumn{1}{c|}{\textbf{F$_{X, 0.3-7keV}$}} & \multicolumn{1}{c|}{\textbf{E$_{s, p}$}} \\

\multicolumn{1}{|c|}{\textbf{}} &
\multicolumn{1}{|c|}{\textbf{}} &
\multicolumn{1}{c|}{\textbf{}} & 
\multicolumn{1}{c|}{\textbf{}} &
\multicolumn{1}{|c|}{\textbf{}} &
\multicolumn{1}{c|}{\textbf{}} &
\multicolumn{1}{c|}{\textbf{$\times$ 10$^{-10}$}} &
\multicolumn{1}{c|}{\textbf{}} \\ 

\multicolumn{1}{|c|}{\textbf{}} &
\multicolumn{1}{|c|}{\textbf{[MJD]}} &
\multicolumn{1}{c|}{\textbf{[s]}} &
\multicolumn{1}{c|}{\textbf{}} & 
\multicolumn{1}{|c|}{\textbf{}} &
\multicolumn{1}{c|}{\textbf{}} &
\multicolumn{1}{c|}{\textbf{[erg cm$^{-2}$ s$^{-1}$]}} &
\multicolumn{1}{c|}{\textbf{ [keV]}} \\
\hline 
\endfirsthead

\multicolumn{8}{c}{{\bfseries \tablename\ \thetable{} -- \astrosat\  data and the best fit parameter corresponding the model : M1}} \\
\hline \multicolumn{1}{|c|}{\textbf{ObsID}} &
\multicolumn{1}{|c|}{\textbf{Time}} &
\multicolumn{1}{c|}{\textbf{Exp.}} &
\multicolumn{1}{c|}{\textbf{\large $\alpha$}} &
\multicolumn{1}{|c|}{\textbf{\large $\beta$}} &
\multicolumn{1}{c|}{\textbf{$\chi^2_\nu$/dof}} &
\multicolumn{1}{c|}{\textbf{F$_{X, 0.3-7keV}$}} &
\multicolumn{1}{c|}{\textbf{E$_{s, p}$}} \\ 

\multicolumn{1}{|c|}{\textbf{}} &
\multicolumn{1}{|c|}{\textbf{}} &
\multicolumn{1}{c|}{\textbf{}} &
\multicolumn{1}{|c|}{\textbf{}} &
\multicolumn{1}{c|}{\textbf{}} & 
\multicolumn{1}{|c|}{\textbf{}} &
\multicolumn{1}{c|}{\textbf{$\times$ 10$^{-10}$}} &
\multicolumn{1}{c|}{\textbf{}} \\ 

\multicolumn{1}{|c|}{\textbf{}} &
\multicolumn{1}{|c|}{\textbf{[MJD]}} &
\multicolumn{1}{c|}{\textbf{[s]}} &
\multicolumn{1}{c|}{\textbf{}} & 
\multicolumn{1}{|c|}{\textbf{}} &
\multicolumn{1}{|c|}{\textbf{}} &
\multicolumn{1}{c|}{\textbf{[erg cm$^{-2}$ s$^{-1}$]}} &
\multicolumn{1}{c|}{\textbf{ [keV]}} \\ \hline 
\endhead

\hline \multicolumn{7}{|r|}{{Continued on next page}} \\ \hline
\endfoot
\hline \hline
\endlastfoot
A02\_199T01\_9000000708 & 57666.25 & 2886.3 & 1.85$\pm$0.02 & 0.45$\pm$0.06 & 267.3/230 & 3.71$\pm$0.06 & 1.46$\pm$0.12 \\ 
A02\_199T01\_9000000708 & 57666.39 & 4041.7 & 1.77$\pm$0.02 & 0.56$\pm$0.05 & 348.9/267 & 3.81$\pm$0.05 & 1.61$\pm$0.1 \\ 
A02\_199T01\_9000000708 & 57666.52 & 2938.6 & 1.81$\pm$0.02 & 0.5$\pm$0.06 & 228.5/238 & 3.94$\pm$0.06 & 1.56$\pm$0.12 \\ 
A02\_199T01\_9000000708 & 57666.6 & 3430.7 & 1.84$\pm$0.02 & 0.45$\pm$0.05 & 280.9/248 & 3.76$\pm$0.05 & 1.51$\pm$0.12 \\ 
G06\_086T01\_9000000774 & 57695.9 & 380.4 & 1.99$\pm$0.07 & 0.29$\pm$0.18 & 46.9/53 & 3.98$\pm$0.19 & 1.03$\pm$0.29 \\ 
G06\_086T01\_9000000774 & 57695.93 & 1286.2 & 1.89$\pm$0.04 & 0.47$\pm$0.08 & 146.2/159 & 3.97$\pm$0.08 & 1.32$\pm$0.13 \\ 
G06\_086T01\_9000000774 & 57696.0 & 924.8 & 1.87$\pm$0.04 & 0.56$\pm$0.1 & 112.1/114 & 3.77$\pm$0.1 & 1.3$\pm$0.13 \\ 
G06\_086T01\_9000000774 & 57696.07 & 577.7 & 1.9$\pm$0.06 & 0.38$\pm$0.14 & 75.1/74 & 3.66$\pm$0.13 & 1.34$\pm$0.28 \\ 
G06\_086T01\_9000000774 & 57696.13 & 233.0 & 1.88$\pm$0.09 & 0.24$\pm$0.21 & 33.4/33 & 4.02$\pm$0.22 & 1.78$\pm$1.17 \\ 
G06\_086T01\_9000000774 & 57696.19 & 706.1 & 1.93$\pm$0.05 & 0.25$\pm$0.12 & 94.5/86 & 3.89$\pm$0.12 & 1.39$\pm$0.39 \\ 
G06\_086T01\_9000000774 & 57696.26 & 370.9 & 1.78$\pm$0.08 & 0.54$\pm$0.18 & 47.1/53 & 3.95$\pm$0.17 & 1.61$\pm$0.38 \\ 
\end{longtable}
\end{center} \begin{center}
\begin{longtable}{|lllllccc|}
\caption[\swift  data and the best fit parameter corresponding the model : M1]{\swift\  data and the best fit parameter corresponding the model : M1. The complete table is available in a machine-readable format in the published journal or via email on request.} \label{tab:swtspec_par} \\

\hline \multicolumn{1}{|c|}{\textbf{ObsID}} & \multicolumn{1}{|c|}{\textbf{Time}} & \multicolumn{1}{c|}{\textbf{Exp.}} & \multicolumn{1}{c|}{\textbf{\large $\alpha$}} & \multicolumn{1}{c|}{\textbf{\large $\beta$}} & \multicolumn{1}{c|}{\textbf{$\chi^2_\nu$/dof}} & \multicolumn{1}{c|}{\textbf{F$_{X, 0.3-7keV}$}} & \multicolumn{1}{c|}{\textbf{E$_{s, p}$}} \\

\multicolumn{1}{|c|}{\textbf{}} &
\multicolumn{1}{|c|}{\textbf{}} &
\multicolumn{1}{c|}{\textbf{}} & 
\multicolumn{1}{c|}{\textbf{}} &
\multicolumn{1}{|c|}{\textbf{}} &
\multicolumn{1}{c|}{\textbf{}} &
\multicolumn{1}{c|}{\textbf{$\times$ 10$^{-10}$}} &
\multicolumn{1}{c|}{\textbf{}} \\ 

\multicolumn{1}{|c|}{\textbf{}} &
\multicolumn{1}{|c|}{\textbf{[MJD]}} &
\multicolumn{1}{c|}{\textbf{[s]}} &
\multicolumn{1}{c|}{\textbf{}} & 
\multicolumn{1}{|c|}{\textbf{}} &
\multicolumn{1}{c|}{\textbf{}} &
\multicolumn{1}{c|}{\textbf{[erg cm$^{-2}$ s$^{-1}$]}} &
\multicolumn{1}{c|}{\textbf{ [keV]}} \\
\hline 
\endfirsthead

\multicolumn{8}{c}{{\bfseries \tablename\ \thetable{} -- \swift\  data and the best fit parameter corresponding the model : M1}} \\
\hline \multicolumn{1}{|c|}{\textbf{ObsID}} &
\multicolumn{1}{|c|}{\textbf{Time}} &
\multicolumn{1}{c|}{\textbf{Exp.}} &
\multicolumn{1}{c|}{\textbf{\large $\alpha$}} &
\multicolumn{1}{|c|}{\textbf{\large $\beta$}} &
\multicolumn{1}{c|}{\textbf{$\chi^2_\nu$/dof}} &
\multicolumn{1}{c|}{\textbf{F$_{X, 0.3-7keV}$}} &
\multicolumn{1}{c|}{\textbf{E$_{s, p}$}} \\ 

\multicolumn{1}{|c|}{\textbf{}} &
\multicolumn{1}{|c|}{\textbf{}} &
\multicolumn{1}{c|}{\textbf{}} &
\multicolumn{1}{|c|}{\textbf{}} &
\multicolumn{1}{c|}{\textbf{}} & 
\multicolumn{1}{|c|}{\textbf{}} &
\multicolumn{1}{c|}{\textbf{$\times$ 10$^{-10}$}} &
\multicolumn{1}{c|}{\textbf{}} \\ 

\multicolumn{1}{|c|}{\textbf{}} &
\multicolumn{1}{|c|}{\textbf{[MJD]}} &
\multicolumn{1}{c|}{\textbf{[s]}} &
\multicolumn{1}{c|}{\textbf{}} & 
\multicolumn{1}{|c|}{\textbf{}} &
\multicolumn{1}{|c|}{\textbf{}} &
\multicolumn{1}{c|}{\textbf{[erg cm$^{-2}$ s$^{-1}$]}} &
\multicolumn{1}{c|}{\textbf{ [keV]}} \\ \hline 
\endhead

\hline \multicolumn{7}{|r|}{{Continued on next page}} \\ \hline
\endfoot
\hline \hline
\endlastfoot
00035025010 & 53884.16 & 3280.4 & 2.01$\pm$0.01 & 0.49$\pm$0.03 & 336.6/307 & 3.26$\pm$0.02 & 0.97$\pm$0.03 \\
00035025118 & 56856.74 & 1276.3 & 1.83$\pm$0.04 & 0.72$\pm$0.09 & 105.1/115 & 1.32$\pm$0.03 & 1.32$\pm$0.09 \\ 
00035025132 & 57052.09 & 1881.8 & 1.94$\pm$0.03 & 0.52$\pm$0.07 & 171.5/163 & 1.29$\pm$0.02 & 1.15$\pm$0.08 \\ 
00035025133 & 57068.11 & 772.1 & 1.91$\pm$0.04 & 0.78$\pm$0.09 & 116.2/104 & 1.94$\pm$0.04 & 1.15$\pm$0.07 \\ 
00035025134 & 57071.54 & 1322.6 & 1.8$\pm$0.03 & 0.87$\pm$0.08 & 167.5/158 & 1.64$\pm$0.03 & 1.3$\pm$0.06 \\ 
00035025135 & 57073.91 & 1437.9 & 2.0$\pm$0.03 & 0.63$\pm$0.07 & 182.6/165 & 1.8$\pm$0.03 & 1.0$\pm$0.05 \\ 
00035025136 & 57080.16 & 1694.1 & 2.08$\pm$0.03 & 0.57$\pm$0.07 & 180.8/164 & 1.47$\pm$0.02 & 0.86$\pm$0.05 \\ 
00035025137 & 57087.9 & 1335.2 & 1.86$\pm$0.02 & 0.64$\pm$0.06 & 226.8/211 & 3.05$\pm$0.04 & 1.29$\pm$0.06 \\ 
00035025139 & 57091.49 & 623.7 & 1.94$\pm$0.03 & 0.62$\pm$0.07 & 177.4/156 & 3.76$\pm$0.07 & 1.12$\pm$0.07 \\ 
00035025141 & 57097.51 & 1919.8 & 2.06$\pm$0.02 & 0.61$\pm$0.06 & 215.9/194 & 3.36$\pm$0.05 & 0.9$\pm$0.04 \\ 
00035025142 & 57099.47 & 1605.1 & 2.06$\pm$0.03 & 0.62$\pm$0.06 & 148.8/176 & 1.92$\pm$0.03 & 0.9$\pm$0.04 \\ 
\end{longtable}
\end{center} 
\end{document}